\documentclass{aa}  

\usepackage{graphicx}
\usepackage{gensymb}
\usepackage{amsmath}
\usepackage{txfonts}
\usepackage{natbib}
\usepackage{graphicx}
\usepackage{tabularx}
\usepackage{subfig}
\usepackage{hyperref}
\begin{document}

\title{The hot dust in the heart of NGC 1068's torus}
\subtitle{A 3D radiative model constrained with GRAVITY/VLTi}


   \author{P. Vermot
          \inst{1}
          \and
          Y. Clénet
          \inst{1}
          \and
          D. Gratadour
          \inst{1}
          \and
          D. Rouan
          \inst{1}
          \and
          L. Grosset
          \inst{2}
          \and
          G. Perrin
          \inst{1}
          \and
          P. Kervella
          \inst{1}
          \and
          T. Paumard
          \inst{1}
          }

   \institute{LESIA, Observatoire de Paris, Université PSL, CNRS, Sorbonne Université, Université de Paris, 5 place Jules Janssen, 92195 Meudon, France \and Kavli Institute for Particle Astrophysics \& Cosmology (KIPAC), Stanford University, Stanford, CA 94305, USA 
             }

   \date{Received September 15, 1996; accepted March 16, 1997}

 
  \abstract
   {The central region of NGC 1068 is one of the closest and most studied Active Galactic Nucleus. It is known to be of type 2, meaning that its accretion disk is obscured by a large amount of dust and gas. The main properties of the obscuring structure are still to be determined.}
   {We aim at modeling the inner edge of this structure, where the hot dust responsible for the near-infrared emission reaches its sublimation temperature.}
   {We use several methods to interpret the K-band interferometric observables from a GRAVITY/VLTI observation of the object. At first, we use simple geometrical models in image reconstructions to determine the main 2D geometrical properties of the source. In a second step, we try to reproduce the observables with K-band images produced by 3D radiative transfer simulations of a heated dusty disk. We explore various parameters to find an optimal solution and a model consistent with all the observables.}
   {The three methods are consistent in their description of the image of the source, an elongated structure with $\sim 4\times6\ mas$ dimensions and its major axis along the northwest/southeast direction. All three of them suggest that of the object looks like an elongated ring rather than an elongated thin disk, with its northeast edge less luminous than the South-West one. The best 3D model is a thick disk with an inner radius $r=0.21_{-0.03}^{+0.02}\ pc$ and a half-opening angle $\alpha_{1/2}=21\pm8\degree$ observed with an inclination $i=44_{-6}^{10}$\,$\degree$ and $PA=150_{-13}^{8}$\,$\degree$. A high density of dust $n=5_{-2.5}^{+5}\ M_{\odot}.pc^{-3}$ is required to explain the contrast between the two edges by self-absorption from the closer one. The overall structure is itself obscured by a large foreground obscuration $A_V \sim 75$. }
   {The hot dust is not responsible for the obscuration of the central engine. The geometry and the orientation of the structure are different from those of the previously observed maser and molecular disks. We conclude that a single disk is unable to account for these differences, and favor a description of the source where multiple rings originating from different clouds are entangled around the central mass. }


   \maketitle
%

\section{Introduction}

In this paper, we present a study of the active galactic nucleus (AGN) of NGC 1068, one of the closest AGN to the Milky Way. Located at a distance of $14.4\ Mpc$ \citep{BlandHawthorn1997}, the spatial scale is $70\ pc/"$ (these conventions will be kept for the whole paper). This proximity, coupled with other observational advantages such as a high luminosity and the absence of foreground emission or absorption makes it a key target in the observation of AGN, and led to a wealth of publications. The observation of its polarised spectrum led to postulate the presence of a dusty torus, and more globally to the unified model of AGN,  proposed by \citep{Antonucci1985, Antonucci1993}.

NGC 1068 nucleus is a complex region, where coexist many structures and various physical conditions. 
From large to small scales, the main components of interest for this study are described hereafter.

The extended region of ionized gas -the Narrow Line Region (NLR)- can be considered as the largest structure of the AGN. It has a characteristic \emph{bicone} or \emph{hourglass} shape oriented along northeast/southwest and motion interpreted as an outflow ejected from the nucleus \citep{Das2006, Poncelet2008, Gratadour2015}. The north pole is oriented towards the observer, with $i=5\degree-10\degree$ and $PA \sim 30\degree$. Two ionization mechanisms are invoked to explain the emission lines properties: photoionization from the accretion disk UV-X radiation \citep{Kraemer2000, Hashimoto2011, Vermot2019} and ionization from shocks due to the interaction between the jet and the interstellar medium \citep{Dopita1996, Exposito2011}.

The most external regions of the torus are cold and not luminous. However, using polarimetric imaging techniques, \citet{Gratadour2015} have detected an elongated $60\times20\ pc$ structure oriented with $PA=118\degree$ tracing scattering of the photons, which is interpreted as the signature of the torus. Another structure, highly polarized, is detected. Its orientation largely differs  from the first, with $PA= 56\degree$. Named \emph{ridge} by the authors, it could arise from dichroic absorption by the dust \citep{Grosset2018, Grosset2019, Grosset2021}.

At smaller spatial scales, thanks to ALMA observations \citet{Gallimore2016, GarciaBurillo2016, Garcia-Burillo2019} highlighted a disk of molecular gas, an elongated $10\times7\ pc$ structure, orientated with $PA = 112\degree$ and $M_{gas} = (1 \pm 0.3) \times 10^5\ M_{\odot}$, with a complex geometry and dynamics. Indeed, in addition to the enhanced turbulence, it has been shown that this disk is counter-rotating \citep{Impellizzeri2019, Imanishi2020} when compared to the inner region where maser spots are detected \citep{Greenhill1996,Greenhill1997, Gallimore2001}.

At similar scales, pieces of information on the warm dust have been gathered by the VLTI/MIDI instrument, with many studies published \citep{Jaffe2004, Weigelt2004, Poncelet2006,Raban2009,Burtscher2013,LopezGonzaga2014}. These studies agree to describe the central source as a $1.4\ pc \times 0.5\ pc$ elongated structure with $PA\sim130\degree - 135\degree$. The temperature of the dust is estimated to be between $600\ K$ and $800\ K$. These studies also revealed the presence and the significant contribution to the flux of a polar emission from colder dust ($T\sim300\ K$), which could originate from the inner edge of the ionization cone \citep{Raban2009}.

VLBA observation revealed even smaller structures, both through continuum \citep{Gallimore1996,Gallimore2001,Gallimore2004} and maser emission \citep{Greenhill1996,Greenhill1997,Gallimore2001}. The source of continuum emission, named S1 at these wavelengths, is resolved with a $0.4\times0.8\ pc$ elongated shape and a major axis oriented along $PA\sim 110\degree$. A detailed analysis of its spectrum in \citet{Gallimore2004} concluded that it results from free-free emission by gas heated at high temperature by the central UV-X source. This lead to a lower limit on its luminosity: $L_{UV-X} \geq 7\times10^{37}\ W$. The detection of $H_2O$ maser spots provides a very reliable and precise measurement of the emitting medium. A line of maser spots, oriented along $PA = 135\degree$ is observed as far as $1.1\ pc$ from the nucleus. The analysis of their velocity profile indicates that they originate from a disk seen edge-on with an inner radius $r_i=0.6\ pc$ and an outer radius $r_o=1.1\ pc$. This maser disk is counter-rotating with the molecular disk observed with ALMA. The maser disk appears contained within the molecular disk and surrounds the cloud of gas from which S1 originates.

A study based on similar GRAVITY data was presented in \citet{GRAVITY2020}, focusing on the interpretation of a reconstructed image of the source and suggesting that the K-band emission originates from a thin ring-like structure with a radius $r=0.24 \pm 0.3\ pc$, $PA=130 \pm \degree$, and inclination $i=70\degree$ with the South Pole pointing toward the observer. In this paper, we present a new analysis of this observation, focusing on the modeling of subset of data with radiative transfer simulations.

\section{Observation}

The GRAVITY observation on which this study is based on was made during the night of November 20th, 2018, under excellent atmospheric conditions. It is one of the two observations used in \citet{GRAVITY2020}.

However, NGC 1068 remains a challenging target to observe with GRAVITY due to its relatively low surface brightness and complex morphology. In order to achieve the observation, $100\%$ of the flux from the source was injected in the fringe-tracker (FT) \citep[see][]{GRAVITY2020}. This strategy made it possible to fringe track, however it means that only FT data are available, with a $R\sim22$ spectral resolution (K-band split in 5 spectral channels) and without any absolute phase measurements. 


Total integration time on the object is 45 minutes, and three reference stars have been observed, HIP 54, HIP 16739, and HIP 17272. Data reduction was done with the standard pipeline provided by ESO. Among others, it provides squared visibility and closure phase measurements. 

\section{Data and interferometric observables}

\subsection{u-v plane}

\begin{figure}
            \includegraphics[width=\hsize]{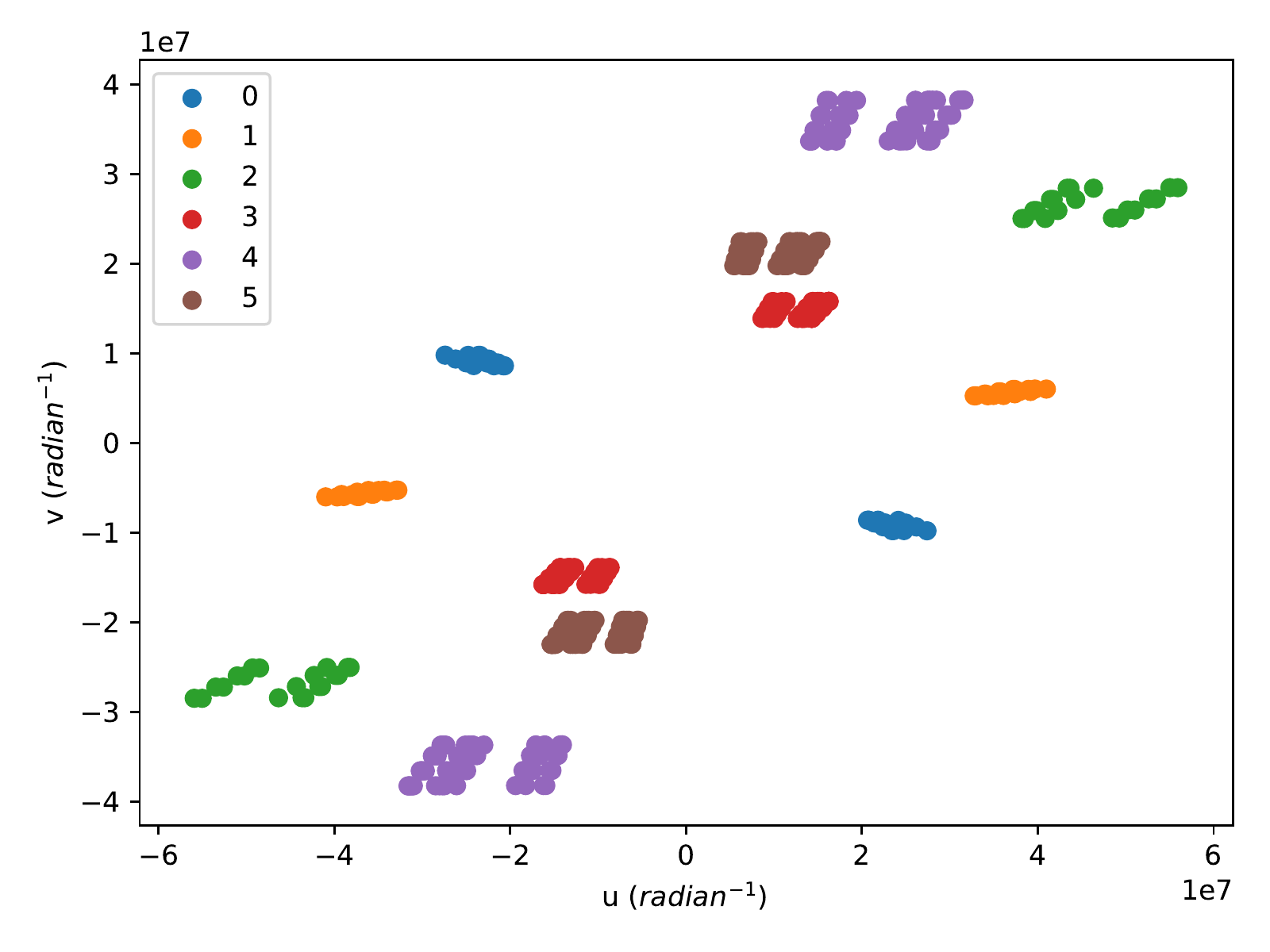}
      \caption{u-v plane associated to the UT observation. (0) UT3-UT4, (1) UT2-UT4, (2) UT1-UT4, (3) UT2-UT3, (4) UT1-UT3, (5) UT1-UT2  
              }
         \label{uv-plane}
   \end{figure}

The spatial frequencies at which the interferometric observables are sampled form the u-v plane of the observation, and are represented in Fig.~\ref{uv-plane}. The sampled spatial frequencies range from $15\times10^6\ \lambda$ to $60\times 10^6\ \lambda$, which probes spatial scales ranging from $3$ to $12\ mas$. The northeast/southwest direction (red, purple, and green bases on Fig.~\ref{uv-plane}) is much more densely sampled than the orthogonal northwest/southeast direction (blue base).

\subsection{Visibility}

For each point of the u-v plane, two visibility estimators are provided by the pipeline, namely the \textit{visibility amplitude} $V_{amp}$ and the \textit{squared visibility} $V^2$. It is this second estimator, less sensitive to phase variations and consequently to atmospheric turbulence, that has been used for this study. Hereafter, the \emph{square root of squared visibility} will simply be named as \emph{visibility}. 
Data selection has been made by removing the visibility points associated with the shortest spectral channel. Their measurement is known to be degraded by the presence of the Gravity metrology laser operating at these wavelengths, as confirmed by the corresponding inconsistent visibility values. 

\begin{figure}
            \includegraphics[width=\hsize]{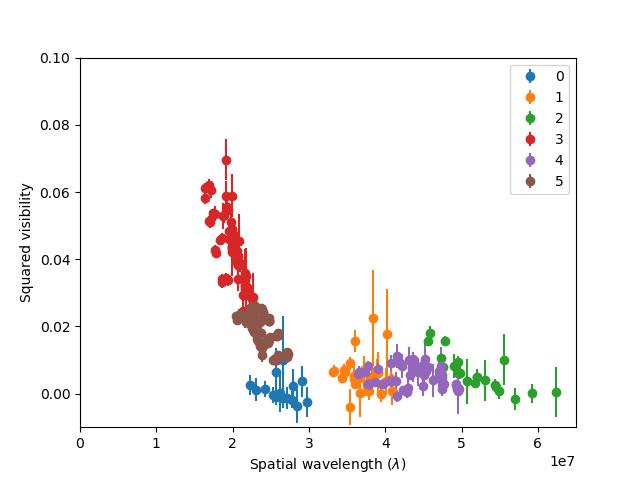}
      \caption{Squared visibility color-coded by baseline : (0) UT3-UT4, (1) UT2-UT4, (2) UT1-UT4, (3) UT2-UT3, (4) UT1-UT3, (5) UT1-UT2  
              }
         \label{vis2}
   \end{figure}

\begin{figure}
            \includegraphics[width=\hsize]{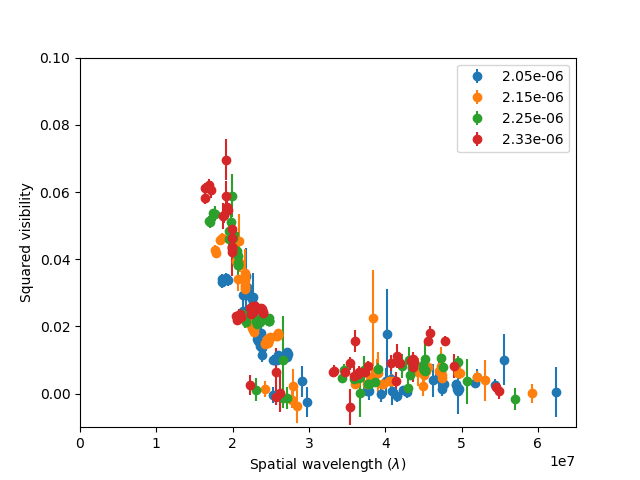}
      \caption{Squared visibility color-coded by wavelength (in meters), as listed in the legend.
              }
         \label{vis2_lam}
   \end{figure}

The visibility points are presented in Fig.~\ref{vis2}, color-coded by baseline. We can notice first a drop in the visibility with increasing spatial frequency up to $25\ M\lambda$ where it reaches zero before slightly increasing at larger frequencies. This indicates that the source is spatially resolved by the interferometer and has a size  between 5 and 20 mas. The \textit{rebound} of the visibility at large frequencies indicates that the main structure exhibits sharp edges. Secondly, the visibility values well below 1 ($\sqrt{V^2}\leq0.25$) reveal that a significant part of the flux comes from a source that can be considered \emph{diffuse} with respect to the resolution and field of view of GRAVITY with the UTs. Fig.~\ref{vis2_lam} represents the same visibility points but color-coded by wavelength. We can notice that the visibility is increasing with wavelength, indicating either that the object's morphology significantly differs from one wavelength to another, or that the ratio of the coherent flux to the total flux increases with wavelength.

The pipeline provides an estimation of the uncertainty associated with the visibility measurement. A detailed study of these estimates revealed that they may not be representative of the actual error on the measurements. Indeed, a significant part of them has $\Delta V^2 < 5 \times 10^{-4}$, well below the scattering of data points observed at the same wavelength and very close spatial frequency. We estimate that this may come from the presence of systematic uncertainties, which are usually not significant compared to the statistical uncertainties but that could become significant in this case. Indeed, the faintness of the source might lead to high systematic uncertainties due to fluctuating AO correction, fiber injection, or fringe jumps. To take into account these systematic errors, we measured the scattering of carefully chosen points at similar wavelength and spatial frequency and decided to add a constant $\Delta V_0 = 0.0015$ uncertainty to every visibility measurement. This allows overcoming problems that occurred with points associated to a very low uncertainty that took an unreasonable importance in the $\chi^2$ fitting procedures, while still taking into account the statistical uncertainties (which for some points can be significant).

\subsection{Closure phase}
\label{subsec:dataset_closurephase}
\begin{figure}
            \includegraphics[width=\hsize]{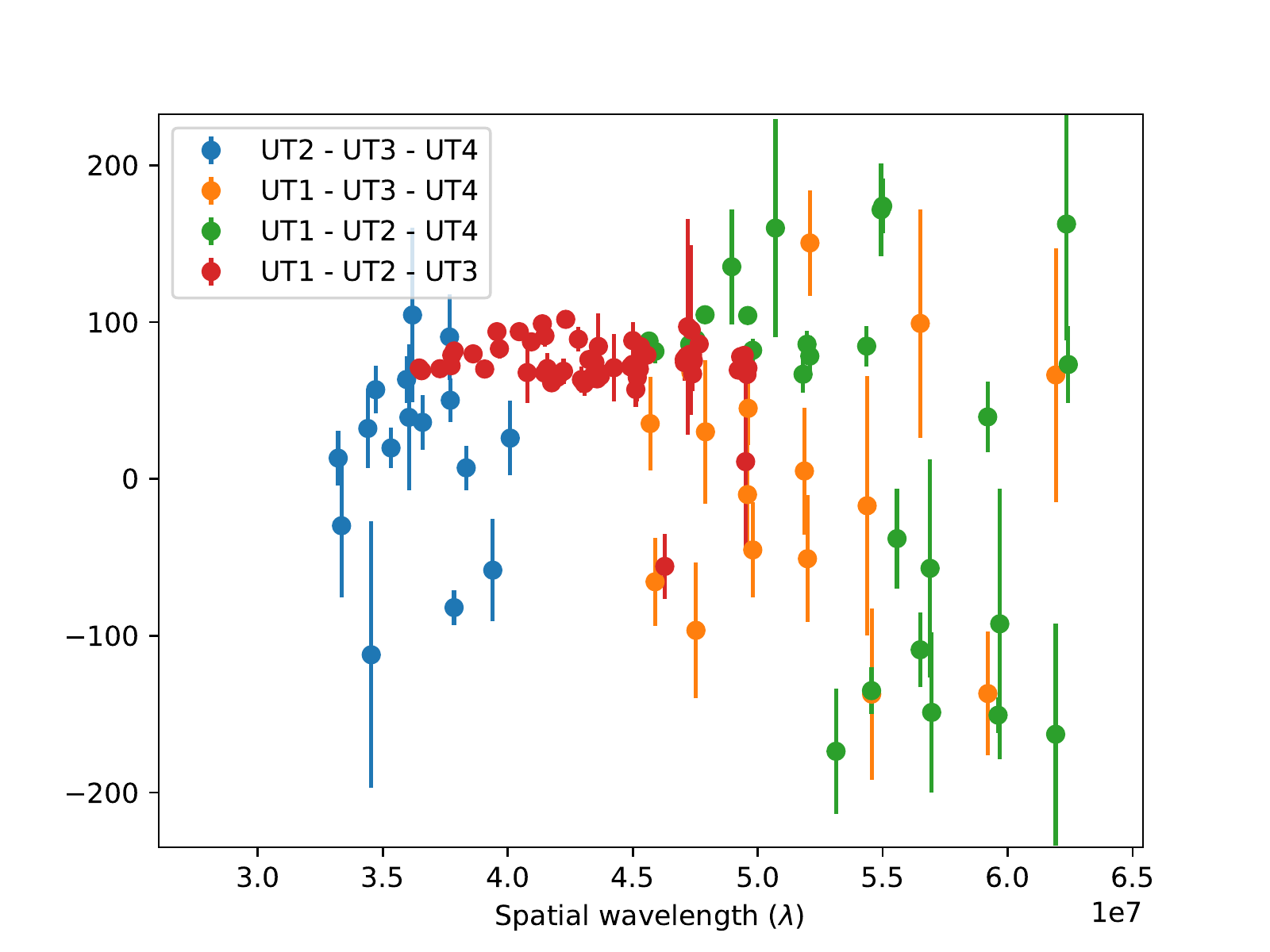}
      \caption{Closure phase color coded by triplet
              }
         \label{t3}
   \end{figure}

The closure phase measurements are presented in Fig.~\ref{t3} and  Table~\ref{table_t3} summarizes some statistical properties of the triplets measurements. 

	\begin{table}[ht]
	\centering
	\caption{\label{table_t3} Basic statistical properties of the closure phase measurements. From left to right: UT telescopes in the triplet, number of closure phase measurements, mean value, median of the uncertainties given by the pipeline, and standard deviation of the measurements.}
	\begin{tabular}{c|c|c|c|c}
	Triplet & N & $\bar{T3} ( \degree)$ & $\tilde{\Delta T3} ( \degree)$ & $std(T3) ( \degree)$ \\
	\hline 
	UT1 - UT2 - UT3 & 60 & 73 & 3 & 19 \\
	UT1 - UT2 - UT4 & 20 & 33 & 10 & 98  \\
	UT1 - UT3 - UT4 & 12 & 52 & 30 & 109 \\
	UT2 - UT3 - UT4 & 12 & -8 & 20  & 21 \\
	\end{tabular}
	\end{table}	
Two remarks can be done from this first statistical analysis. First, over 104 closure phases measurements, 60 are provided by the [UT1-UT2-UT3] triplet. Moreover, the uncertainties estimated on the points coming from this triplet are much smaller than those from the three others: the median of the uncertainties is $3\degree$ for [UT1-UT2-UT3] while it is $21\degree$ for the 44 other measurements. In consequence, in the upcoming analysis, this triplet will have a larger weight in the data fitting involving the closure phase measurements. Second, the closure phase mean value for this triplet differs from 0 ($\bar{T3} = 73\degree \pm 19\degree$), which already indicates that a significant asymmetry is present in the source luminosity distribution.
   
\subsection{Coherent spectrum and magnitude}

\begin{figure}
            \includegraphics[width=\hsize]{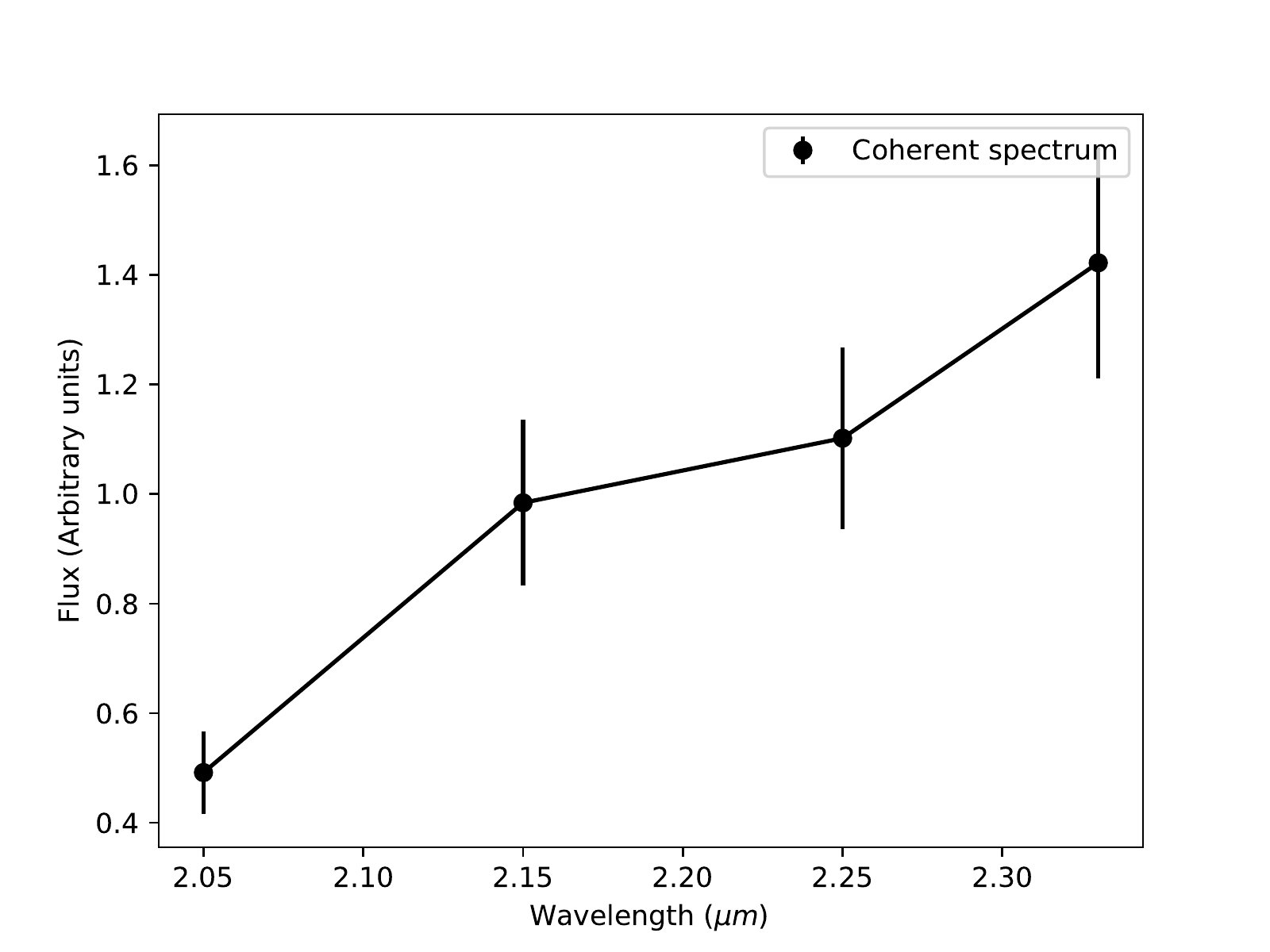}
      \caption{Observed coherent spectrum.
              }
         \label{spec_cohe}
   \end{figure}

The coherent flux measurements provided by GRAVITY are presenting a very large scattering, most probably due to the efficiency in the fiber injection which varies during the observation. Hence, it cannot be used alone, and we computed instead the mean of this observable over all spatial frequencies, to obtain an estimation of the spectrum of the resolved source. From now on, this spectrum will be called \emph{coherent spectrum}. It is the 4-point low-resolution spectrum ($R\sim22$) in arbitrary units presented in Fig.~\ref{spec_cohe}. We can notice that the flux increases with wavelength, and that a slight excess is present at $2.15\ \mu m$. Considering the uncertainties associated to this spectrum, this could be due to random fluctuation. However, we note that this could also indicate the presence of a strong Brackett $\gamma$ emission line.

At last, the detailed analysis of the images from the acquisition camera presented in \citet{GRAVITY2020} provides an estimation of the total K-band magnitude injected in the fiber: $K_{tot}=8.33\pm0.25$. 

\subsection{AT observation}

We decided not to include in our analysis another observation which was performed with the Auxiliary Telescopes (AT) of the VLTI. The motivation for this decision is twofold.

First, it is difficult to merge the visibility measurments from the UT and AT observations. Indeed, at least two incompatibilities appear in the visibility measurements done on adjacent and over-imposed baselines from the two observations.

We consider that these differences appear because of the different relative contributions of the diffuse background to the AT and UT fields of view. Indeed, the fibers of the UT have a $\sim$60$\ mas$ FoV while the AT have a $240\ mas$ FoV. We will see in the following sections, as already stated in \citet{GRAVITY2020} and as suggested by the MIDI observations, that the diffuse background contributes significantly to the total flux injected in the UT FoV. In consequence, the diffuse flux in the FoV of the AT is likely larger than the UT, while the flux from the resolved structure will remain identical. In that case, if these effects are not included in the model these jumps in the visibility measurement might be wrongly interpreted as resulting from a feature of the resolved source. 

A model of these extended structures would be required to take into account their respective effect on each FoV. This is especially true since this effect appears to vary according to the direction of the baseline. In the upcoming analysis, we assume that the diffuse contribution is uniform. While it might be sufficient for the very small scales that we are modeling, such an assumption cannot be valid at the spatial scale of the AT FoV. 

The second motivation to discard the AT observations was to focus on the smaller spatial scales probed with the UT.

\section{Modeling}
\subsection{Geometrical approach}
\label{geom_models}

As a first step, we used geometrical approaches to model the observables and derive images of the source: we used simple geometrical models fitted with the LITpro software \footnote{LITpro software available at http://www.jmmc.fr/litpro} as well as image reconstructions using the software MiRa \citep{Thiebaut2003}. The goal was to look for characteristic size measurements and hints on the morphology to guide our further radiative transfer modeling.

All simple geometrical models converge toward a central elongated structure (see Table \ref{fit_geom}), with $\sim 4\times6\ mas$ ($\sim 0.28\ \times 0.42\ pc$) dimensions and a major axis at $PA\sim140\degree$. Also, all models agree on the significance of the diffuse flux, contributing to around $2/3$ of the total flux. Out of the four simple models presented in Table \ref{fit_geom}, the one which provides the best fit to the data is the elongated thick ring, whose K-band image image and associated visibility points are presented in Fig. 6.

As in \citet{GRAVITY2020}, we conclude from our simple image reconstruction attempts on the presence of an inclined ring or disk with $\sim 3.5\ mas\ (0.25\ pc)$ radius, an elongation ratio $L/l\sim0.5$ and the \textit{northeast} edge less luminous. Still, we observe significant variations of the aspect of the source from one reconstruction to another, especially along the poorly sampled Nort-West/South-East direction, which prevent us from reliably constrain a model on the images. 

We can also notice that this first modeling is consistent with results from VINCI \citep{Wittkowski2004} or MIDI observation \citep{Jaffe2004,Honig2008,Raban2009,LopezGonzaga2014} that revealed a structure with a very similar orientation. These latter studies however found a larger size for the structure, which is explainable by the differences in temperatures probed by the two instruments.



	
	\begin{table*}[ht]
	\centering
	\caption[Results for the different geometrical models]{\label{fit_geom} Results of the fit performed for each geometrical model.}
	\begin{tabular}{c|c|c|c|c}
	 & Gaussian & Disk & Thin ring & Thick ring \\
	\hline
	Minor axis & $4.02 \pm 0.10$ mas & $ 5.00 \pm 0.17 $ mas & $3.73 \pm 0.05$ mas & $3.51 \pm 0,04$ mas  \\
	Elongation ratio & $1.55 \pm 0.6 $ & $ 1.76 \pm 0.07 $ &  $1.50 \pm 0.04 $ & $1.51 \pm 0.04 $ \\
	PA & $140.72 \pm 2.42 \degree $ & $ 139.73 \pm 1.56 \degree $  & $ 144.08 \pm 1.98 \degree $ & $ 138.07 \pm 1.62 \degree $ \\
	Background flux & $ 1.81 \pm 0.07 $ & $2.40\pm 0.08 $ & $2.17\pm 0.04 $ & $2.04 \pm 0.03$   \\
	\hline
	$\chi^2$ & $3.89$ & $5.10$ & $4.03$ & $3.47$ \\
	\end{tabular}
	\end{table*}

	\begin{figure*}[!ht]
	\centering		
	\subfloat[c][]{\includegraphics[width=0.4\hsize]{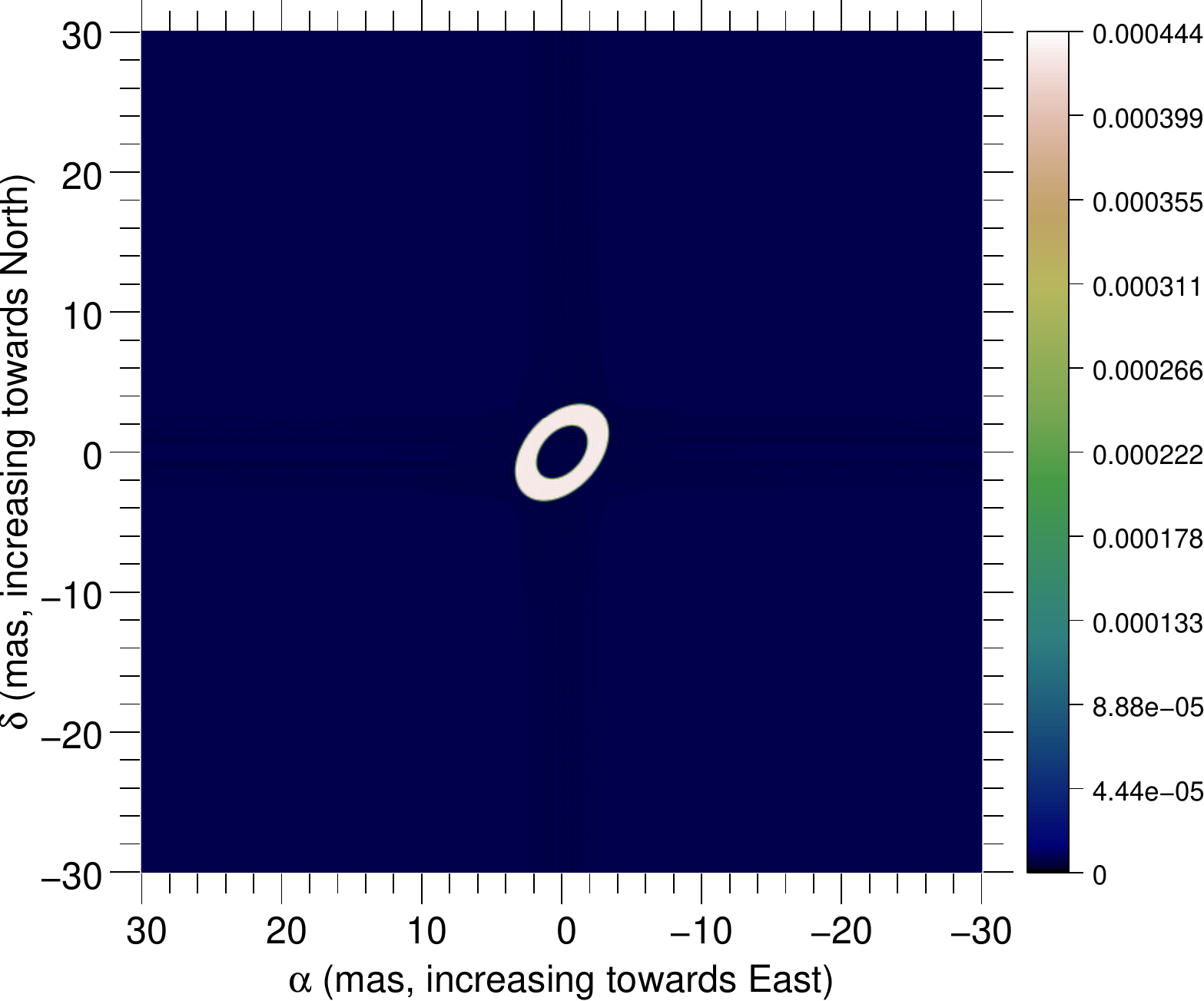}\label{im_thickring}}
	\subfloat[c][]{\includegraphics[width=0.4\hsize]{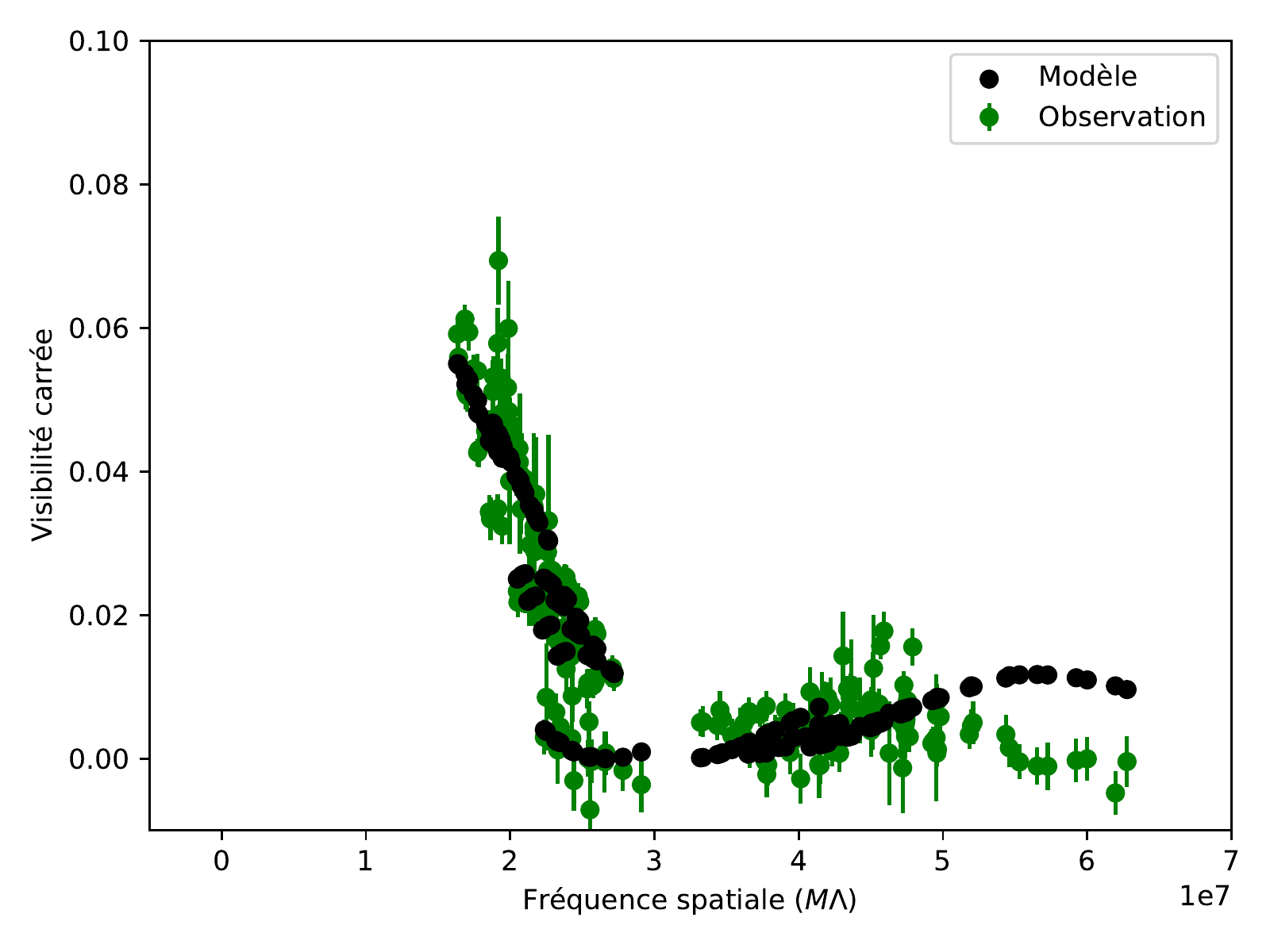}\label{vis_thickring}}
      \caption{\protect\subref{im_thickring} Image in arbitrary flux units of the best geometrical model, an elongated thick ring over-imposed to a diffuse emission. \protect\subref{vis_thickring} Comparison between the observed and the modeled visibilities
              }
              \label{imvis_thickring}
	
	\end{figure*}

\subsection{Physical modeling from radiative transfer simulations MontAGN}

In this section, we describe how we used radiative transfer simulations to derive a physically realistic 3D model of the source from the GRAVITY observables.


We are trying here a new approach, using a model highly constrained from astrophysical considerations in order to produce physically realistic images, that will be then compared to the interferometric measurements in a second step. For this purpose, we are using the simulation code MontAGN \citep{Grosset2018, Grosset2019} to model a dusty disk heated by a central source of energetic photons. The images produced by each model are used to compute the various interferometric observables. 

We first give a description of MontAGN, and general considerations on the model that will be used. We then describe two sets of simulations, the first one with sole purpose to reproduce the visibility and the photometric measurements, the second one especially designed to additionally take into account the closure phase measurements.

\subsubsection{MontAGN}

MontAGN is a radiative transfer simulation code using Monte Carlo methods, developed to study AGN dusty tori, and more specifically the interpretation of polarimetric observations. It offers the possibility to model the diffusion, absorption, and emission of photons by dusty structures with arbitrary geometries for a wide range of dust types.

MontAGN follows the path of photons, from their emission by a central source until the moment they exit the simulation grid, modeling all along the different interactions with the medium.

In the following list, we succinctly describe the main steps of interest of a simulation:
\begin{enumerate}
\item Extinction and absorption coefficients, albedo, Mueller matrices, and phase functions are calculated on a wide range of wavelengths (from X to far-IR) and grain sizes for each of the "dust types" defined by the user
\item The 3D grid of cell is initialized with Cartesian coordinates, and is filled with dust by attributing a density value for each cell and each dust type.
\item Photons are emitted by the central source with a random direction, by monochromatic packets, with an initial wavelength randomly chosen according to the source SED.
\item Each packet of photons propagates freely in a straight-line through empty cells. When the packet enters a non-empty cell, densities are used to randomly decide if the photon will interact with the cell. If it does not, it pursues its straight line propagation.
\item If the photon interacts with the cell, the type and size of the grain with which it will interact, then the type of interaction (absorption or diffusion), are randomly chosen according to the densities, extinction coefficients, and albedo of the different grains present in the cell.
\item If the photon is scattered, its wavelength stays unchanged, its new direction is chosen from the phase function of the selected grain, and its Stokes vector is updated. If the photon is absorbed, the temperature of the cell is updated, and a new packet of photons with the same energy is emitted in a new direction, with a wavelength randomly chosen to match the Planck emission of the cell.
\item Eventually the photons exit the simulation grid. Their properties are then stored in an output file.
\item From these files containing information on every photon exiting the simulation grid, images can be generated for any position of the observer and any spectral domain.

\end{enumerate} 
\subsubsection{Base model}

   \begin{figure}
            \includegraphics[width=\hsize]{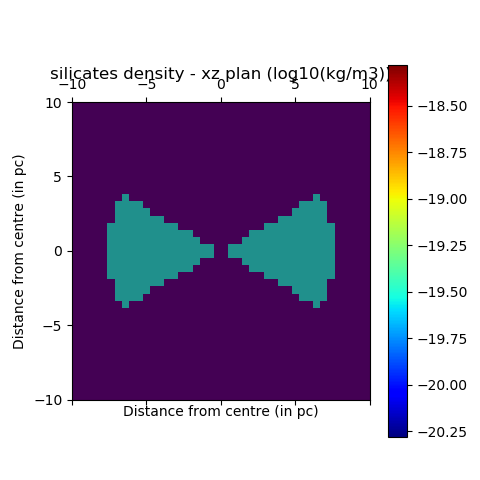}
      \caption{Geometry of the thick disk model used in the MontAGN simulations. The image is a cut along any direction orthogonal to the equatorial plane of the structure.
              }
         \label{geometry}
   \end{figure}
   \begin{figure}
            \includegraphics[width=\hsize]{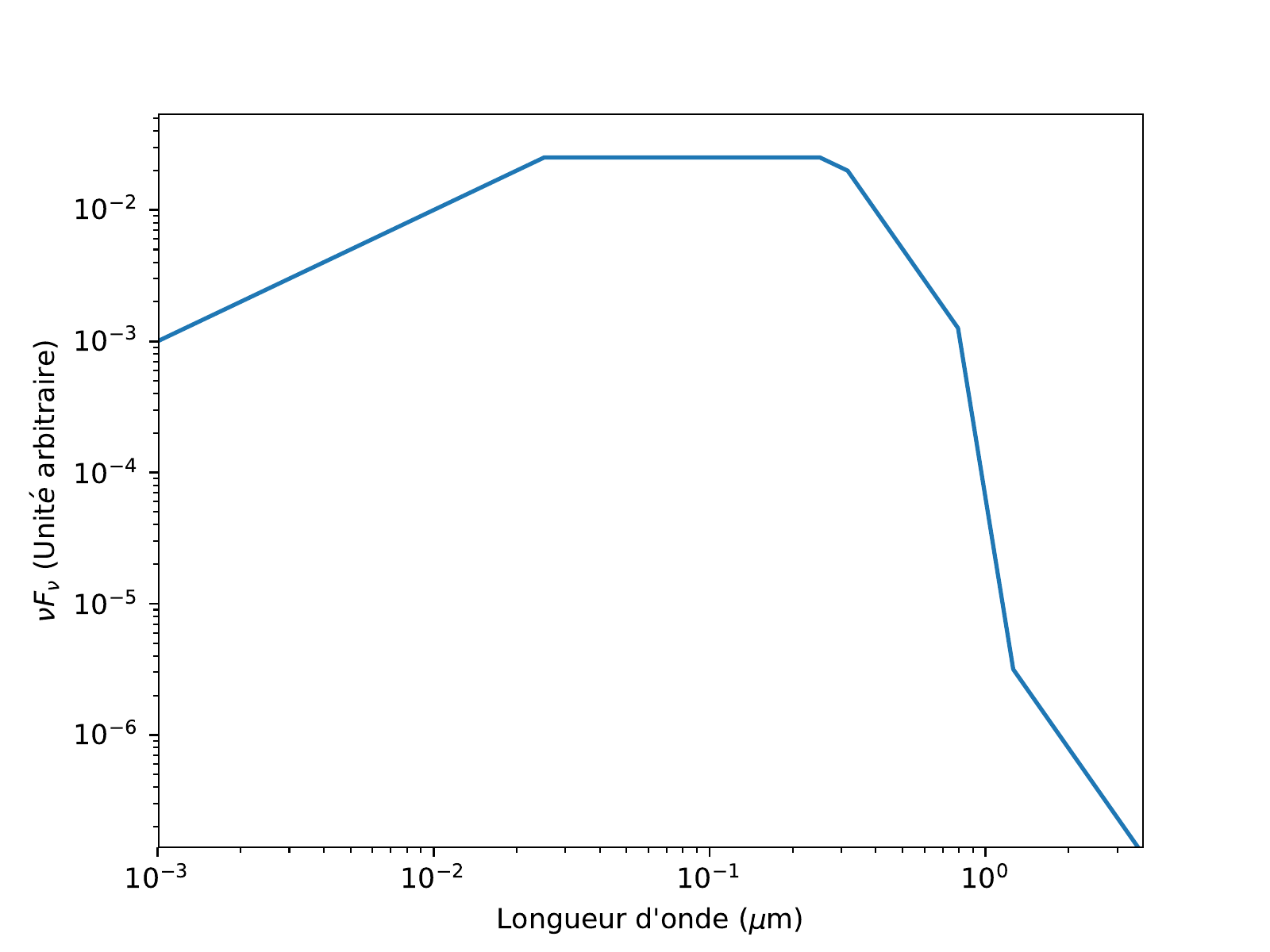}
      \caption{Spectrum of the central source used in  MontAGN simulations
              }
         \label{spectrum}
   \end{figure}

We model the hot dust of the inner region of NGC 1068's dusty torus by graphite grains, distributed up to sublimation limit in a thick disk-like structure (see Fig. \ref{geometry}) with a uniform density. Guided by the size estimations from geometrical models and image reconstructions, the grid has been chosen to be contained in a 1 pc wide cube with $0.025\ pc$ cells.
   
Interstellar dust mostly contains two grain types, respectively graphite and silicate \citep{Barvainis1987, Netzer2015} which are simultaneously present in most astrophysical conditions. However, close to the accretion disk of an AGN, dust grains are heated by the UV-X emission until they sublimate. Two distinct arguments point to the conclusion that silicate grains will reach sublimation at a larger distance from the accretion disk than graphite grains, which would then be the sole one populating this inner region of the dusty tori. First, silicate grains dispel their thermal energy in the infrared less efficiently than graphite ones, while they are absorbing similarly UV-X photons. So, at a given distance from the central source, silicate grains will be found at higher temperatures. Secondly, the sublimation temperature of silicate grains is lower than the one of graphite grains \citep[][$T_{sub, silicates}\sim1400\ K, T_{sub, graphites}\sim1700\ K$]{Barvainis1987, Baskin2018}. Hence, our model only contains graphite grains and we make the hypothesis that the inner radius of the dusty structure coincides with the distance at which they sublimate. The size distribution of the grains ($n$ grains of size $a$) is assumed to follow a MRN power-law \citep{Mathis1977}) with $dn/da \propto a^{-3.5}$ and $a_{min} = 5\ nm$ and $a_{max} = 250\ nm$. Given this size distribution, the mean density of grains specified as an input of the model, and the geometry of the structure, we can compute the total mass of dust of the model.

To describe the geometry of the hot dust structure, we use a simple model, with a geometrically thick disk defined by its inner radius,  outer radius, and  opening angle (see Fig.~\ref{geometry}). As we will see in the following section with the images generated by MontAGN, the luminosity of the dust decreases very quickly with the distance to the central source and most of the K-band emission comes from a region very close to the sublimation distance, forming a thick ring. Hence, the large scale structure of the torus has only little impact on the image of the object at these wavelengths and a more complex model is not required at this stage. The opening angle of the structure (i.e., the thickness of the disk) will be a variable parameter, as well as its inner radius. The outer limit is defined by the grid limits.

At last, since the accretion disk of NGC 1068 is obscured and that little information is known about its actual spectrum, we are using to describe the central source (considered as point-like) a simple spectrum, similar to the one used in \citet{Honig2006} (see Fig.~\ref{spectrum}). It emits the vast majority of its flux in the UV-X domain, with a negligible infrared luminosity. Note that for verification, a few simulations have been run with other central spectra, without any detectable effect as long as most of the energy is emitted in the UV-X domain. For a given type of grain and a given central source spectrum, the sublimation radius is only determined by the luminosity of the source. 

\subsubsection{MontAGN model 1: accounting for visibilities, photometry and spectral variations}

\begin{table}[ht]
	\centering		
    \begin{tabular}{l|c}
      Parameter & Fixed value or \\
       & [min. value, max. value, sampling] \\
      \hline
      Dust type & Graphite \\
      Minimum grain size & $5\ nm$ \\
      Maximum grain size & $250\ nm$ \\
      Power law index & $-3.5$  \\
      Density & $2\ cm^{-3}$ \\
      \hline
      Inner radius & [$0.15, 0.35, 0.01$] pc \\
      Half-opening angle & [$5, 39, 2$]$\degree$ \\
      \hline
      Inclination & [$30, 90, 1$]$\degree$ \\
      PA & [$0, 360, 1$]$\degree$ \\
      \hline
      Field of view & $2.5\ pc$ \\
      Pixel size & $0.025\ pc$ \\
      \hline
      Number of photons & $4 \times 10^6$ \\
    \end{tabular}
    \caption{\label{param_modele_12} MontAGN simulation parameters for model 1}
\end{table}

	\begin{table}[ht]
	\centering		
    \begin{tabular}{l|c}
      Parameter & Value \\
      \hline
      & \\
      Inner radius & $0.23_{-0.04}^{+0.02}$ pc \\
      & \\
      Half-opening angle & $9_{-4}^{+16}$ $\degree$ \\
      & \\
      Inclination & $52_{-4}^{+2}$ $\degree$ \\
      & \\
      PA &  $133_{-15}^{+15}$ $\degree \pm 180\degree$ \\
      & \\
	  $A_V$ & $70 \pm 0,2$\\
    \end{tabular}
    \caption{\label{model_1} Best fit parameters from model 1. Uncertainties are estimated with $\chi^2 \leq 2\chi^{2}_{min}$. $PA$ is defined modulo $180\degree$, see text.}
\end{table}
	
For the first model, we only focus on reproducing the visibility points without taking into account the closure phase, to constrain the geometry of the source and the main scaling parameters of the model. More precisely, we explore the values of the sublimation radius, the opening angle, the inclination, and polar angle (PA) parameters as reported in Table \ref{param_modele_12}.

Once a simulation with a given sublimation radius and opening angle is finished, K-band images with a $2.5\ pc\times 2.5 \ pc$ field of view and $0.025\ pc$ pixel size are generated for a range of inclinations. The discrete Fourier Transforms are computed and linearly interpolated at the spatial frequencies observed with GRAVITY. PA values are explored by a change of coordinates so that fewer images have to be computed.



The spectrum of the model never matches the observed coherent spectrum: it is too \emph{blue}, even in the most favorable cases. Moreover, the analysis of the simulation reveals that the models are too luminous when compared to the K-band magnitude. These two remarks suggest that the infrared emission of this region is significantly absorbed by foreground material.
To match the observed K-band magnitude, we fit for each set of parameters a standard extinction \citep{Cardelli1989} by applying its effect uniformly on the initial MontAGN images. The $A_V$ values reported in Tables \ref{model_1} and \ref{model_2} correspond to this extinction, which is not caused by the hot dusty structure, but rather by foreground colder material. Then, we deduce the spectrum of the diffuse component required to match the observed visibility.

Table \ref{model_1} gives the value of the best fit parameters and Fig.~\ref{chi2_1} of the appendix shows cuts around this best solution in the 4D $\chi^2$ cube. We can notice that the inner radius, the inclination, and the position are well constrained, with well-defined minima. However, two solutions are possible for PA, separated by exactly $180\degree$ which highlights the limitations of a modeling performed solely on visibility measurements.

The opening angle of the disk is not well constrained, and even if an optimal solution is obtained, the $\chi^2$ analysis reveals that we can only reliably consider an upper limit on this parameter: $\alpha_{1/2} \leq 25 \degree$. 

From Fig.~\ref{chi2_1} in the appendix, we can notice that one slight degeneracy seems to appear between the inclination and the sublimation radius. It is explained by the fact that given the u-v plane of the observation, the information contained in the visibility is mostly a measurement of the spatial extent of the structure in the northeast/southwest direction.

Figure~\ref{im_source_mod1} shows the K-band image associated to this best solution. Because of the $52_{-4}^{+2}$\,$\degree$ inclination, the source looks like an inclined ring. The low density of dust used in this model explains both the symmetric aspect of the structure and the fact that the northwest and southeast extremities are more luminous since they correspond to line of sights geometrically crossing more hot dust.

These results are in good agreement with the ones from the geometric models presented in Sect.~\ref{geom_models} and in \citet{GRAVITY2020}, indicating in first instance that the K-band emitting structure looks like an inclined ring with $PA\sim 135\degree$.

\subsubsection{MontAGN model 2 : accounting also for closure phase}   
\label{sec:model1}
   	
\begin{table}[ht]
	\centering		
    \begin{tabular}{l|c}
       & Fixed value or \\
      Parameter & [value min, value max, step] or \\
       & {values}\\
      \hline
      Dust type & Graphite \\
      Minimum grain size & 5 nm \\
      Maximum grain size & 250 nm \\
      Power-law index & -3,5  \\
      Density & \{1; 1.5; 2; 2.5; 3; 3.5; 4; 4.5; \\
      & 5; 10; 20; 30; 40\} $cm^{-3}$ \\
      
      \hline
      Inner radius & [0.15, 0.3, 0.01] pc \\
      Half-opening angle & [5, 33, 4]$\degree$ \\
      \hline
      Inclination & [30, 90, 1]$\degree$ \\
      PA & [0, 360, 1]$\degree$ \\
      \hline
      Field of view & 2.5 pc \\
      Pixel size & 0.025 pc \\
      \hline
      Number of photons & $2 \times 10^6$ \\
    \end{tabular}
    \caption{\label{param_modele_2} MontAGN simulation parameters for model 2}
\end{table}

	\begin{table}[ht]
	\centering		
    \begin{tabular}{l|c}
      Parameter & Value \\
      \hline
      Inner radius & $0.21_{-0.03}^{+0.02}$ pc \\
      Half-opening angle & $21_{-8}^{+8}$ $\degree$ \\
      Inclination & $44_{-6}^{+10}$ $\degree$ \\
      Density & $10_{-5}^{+10}\ cm^{-3}$ \\
      PA &  $150_{-13}^{+8}$\,$\degree$ \\
	  $A_v$ & $76.5 \pm 0.3$\\
    \end{tabular}
    \caption{\label{model_2}Best fit parameters for model 2. Uncertainties are estimated with $\chi^2 \leq 2\chi^{2}_{min}$.}
\end{table}
	
\begin{figure*}
\centering
            \includegraphics[width=0.8\hsize]{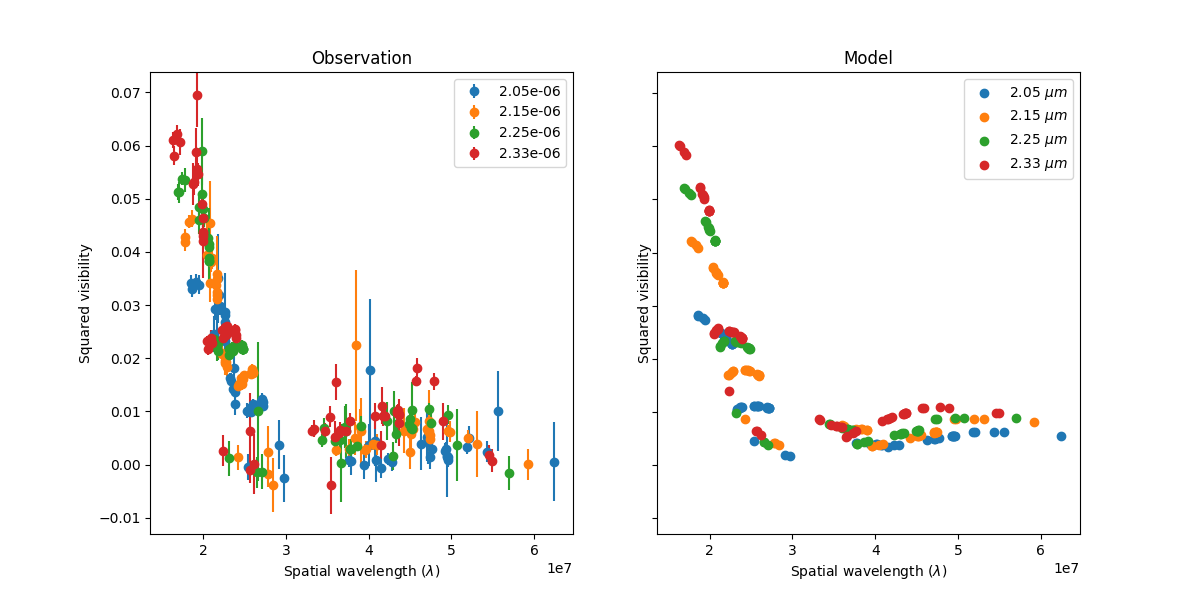}
      \caption{Comparison between the observed visibility (left) and the one predicted by the MontAGN model  2 (right)
              }
              \label{vis_mod2}
   \end{figure*}

\begin{figure*}
\centering
            \includegraphics[width=0.8\hsize]{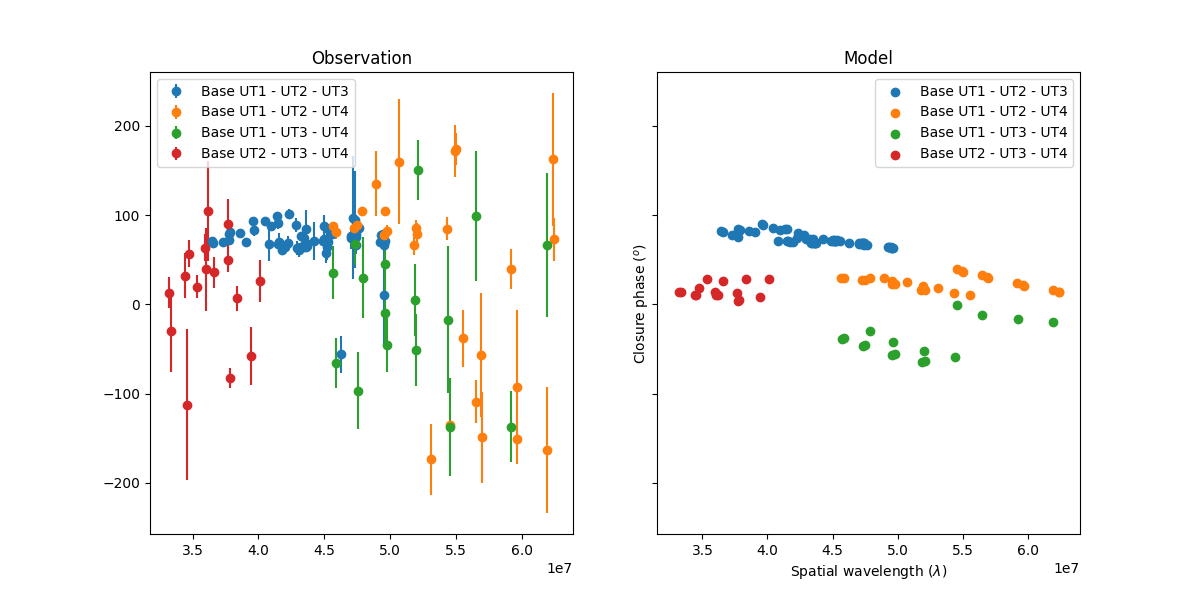}
      \caption{Comparison between the observed closure phase (left) and the one predicted by the MontAGN model 2 (right)
              \label{t3_mod2}
              }
   \end{figure*}

	\begin{figure*}[!ht]
	\centering		
	\includegraphics[width=0.7\hsize]{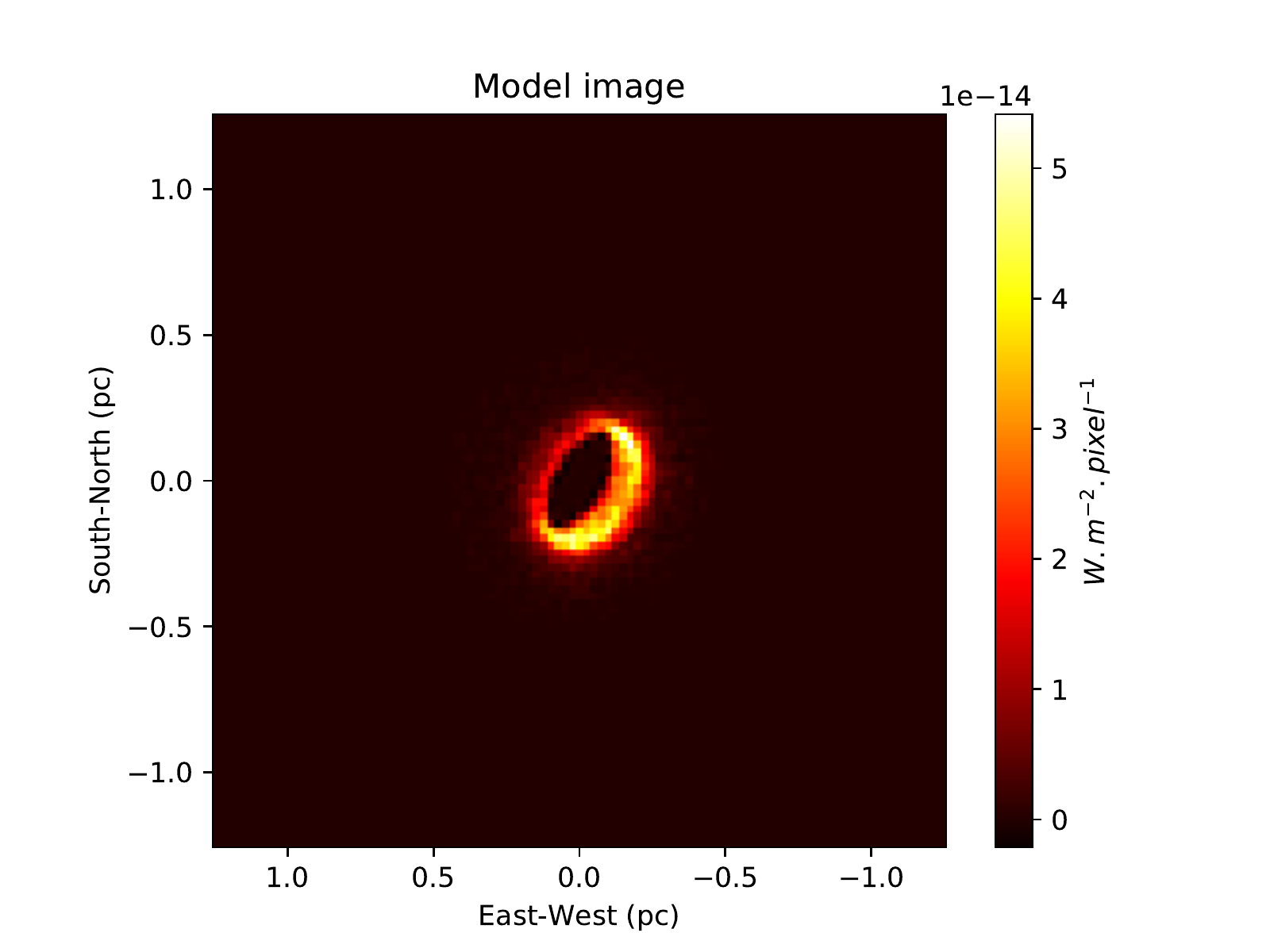}
	\caption{K-band image of the MontAGN model 2. The diffuse background is not represented.}
	\label{im_source_mod2}
	\end{figure*}

In a third model, we try to reproduce simultaneously the GRAVITY closure phase and visibility measurements.

As already pointed out in Sect.~\ref{subsec:dataset_closurephase}, the [UT1-UT2-UT3] triplet (in red in Fig.~\ref{t3}) contains the majority of the closure phase measurements and is moreover associated with the lowest uncertainties. The spatial frequencies probed by this triplet allow to constrain an important property of the model: the difference of luminosity between the northeast and the southwest regions of the structure. The closure phase is positive when the southwest edge is the most luminous, and negative in the other case. Their value is an indicator of the relative contrast between the two edges. The mean closure phase on the [UT1-UT2-UT3] triplet is $73\degree$, which implies a high contrast.

The first dataset of MontAGN simulations, which was used for model 1 in Sect.~\ref{sec:model1}, produces very symmetric images which do not allow to reproduce this contrast and the closure phase measurements. However, increasing the dust density of the disk in MontAGN simulations results in an increase of the contrast between the two edges: the closest edge becomes optically thick,  self-absorbs the infrared photons it emits and as a result, its K-band luminosity is decreased. Hence, the closure phase of the resulting model for the [UT1-UT2-UT3] triplet gets closer to the observed ones.

In order to investigate this new aspect of the model, new simulations are performed, with a coarser sampling for the opening angle (accounting for the difficulty to constrain this parameter and hence to spare time) and with a wide range of dust densities (testing up to 40 grains per $cm^3$ was necessary to converge toward a solution).

The comparison with the visibility measurements is similar to the one performed for model 1. The complex discrete Fourier Transform of the images being already computed and interpolated at the spatial frequency probed by the baselines, the computation of the closure phase is direct. The presence of a diffuse background, a foreground extinction, and the luminosity of the source does not influence these closure phase measurements, so there is no additional correction to apply.

Two 5D $\chi^2$ maps are computed, one for the visibility predictions of the model and one for the closure phase. After normalization, done so that the minimum of each hypercube equals the number of data points of the associated observable, they are added to get the final $\chi^2$ estimator.

The various 2D $\chi^2$ maps are presented in Figs.~\ref{chi2_2a} and~\ref{chi2_2b}. It is interesting to note that there is a slight degeneracy between the opening angle and the inclination, maintaining the \emph{annular} aspect of the source. Nevertheless, these maps highlight that the existence of a well defined and unique solution in the explored parameter space.

The best model parameters are presented in Table~\ref{model_2}. Some properties from previous models are maintained, but, in addition to providing a constraint on the dust density, the inclusion of the closure phase impacted some of the geometrical parameters. 

A comparison between the observed and the predicted visibility is presented in Fig.~\ref{vis_mod2}, and the comparison of the closure phase in Fig.~\ref{t3_mod2}. The prediction of the visibility is similar to the one obtained with model 1; the low spatial frequencies are well reproduced with their spatial variability, the rebound at high frequencies also, even if the spectral dispersion is underestimated by the model at these frequencies. Except the shortest wavelength of the shortest baseline, this model reproduces well the observed visibility and its spectral dependency.

The predictions of the model agree correctly with the observed closure phase measurements, more particularly those related to [UT1-UT2-UT3] triplet (blue), which carries most of the information. The [UT2-UT3-UT4] triplet (red) is also fairly well reproduced by the model,  as well as for the [UT1-UT3-UT4] triplet (green) considering the uncertainties. However, the [UT1-UT2-UT4] triplet (orange), which probes the highest spatial frequencies, is not well reproduced by the model. The uncertainties on the values of this triplet and their scattering are high, which clearly indicates the lack of reliability of this measurement. However, closure phase measurements are known to be very robust and we consider that a better explanation is that this observation traces an additional asymmetry at the smallest spatial scales, which cannot be reproduced by the models used in this work.

With $i=44\degree$ and $\alpha_{1/2}=21\degree$, the structure still looks like an inclined ring. However, its orientation is now estimated to be $150_{-13}^{+8}$\,$\degree$, against $\sim 135-145\degree$ for previous models. Considering the estimated uncertainties, these values could be compatible, but the difference is significant. The closure phase provides an estimation of the density of the medium: $n_{grains}=10\ cm^{-3}$. At last, the temperature of the diffuse component is estimated to be $T \sim 600\ K$ and the foreground extinction to be $A_V = 76.5$ (i.e., $ A_K = 8.9$), very similar to the one deduced from the previous model.

The best K-band image of the source is presented in Fig.~\ref{im_source_mod2}. It is compatible with our previous models in terms of shape and size, despite noticeable differences. The most significant one is of course the high contrast between the two edges of the structure. It can be explained as follows: for the South-West edge, we have a direct line of sight toward the surface of sublimation, while for the northeast edge it is obscured by the dust located in the line-of-sight, so that only the South extremity of this edge can be observed. This last point may explain that for a similar aspect of the image, this model corresponds to a lower inclination.

Hence, knowing that the northeast edge is the less luminous allows us to state that the \emph{South Pole} of the structure is directed toward the observer.

	\begin{figure*}[!ht]
	\centering		
	\subfloat[][]{\includegraphics[width=0.49\hsize]{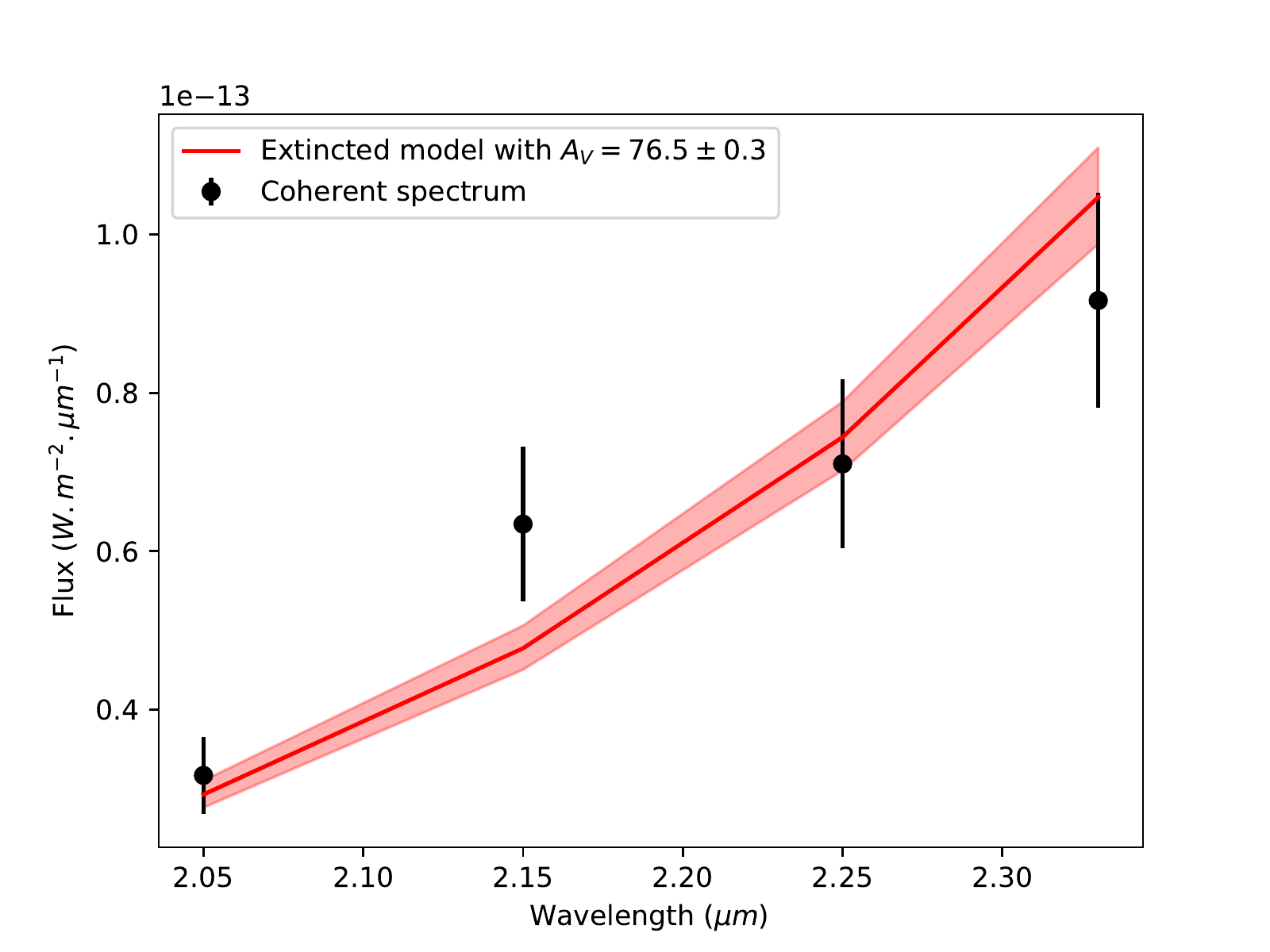}\label{spec_cohe_2}}
	\subfloat[][]{\includegraphics[width=0.49\hsize]{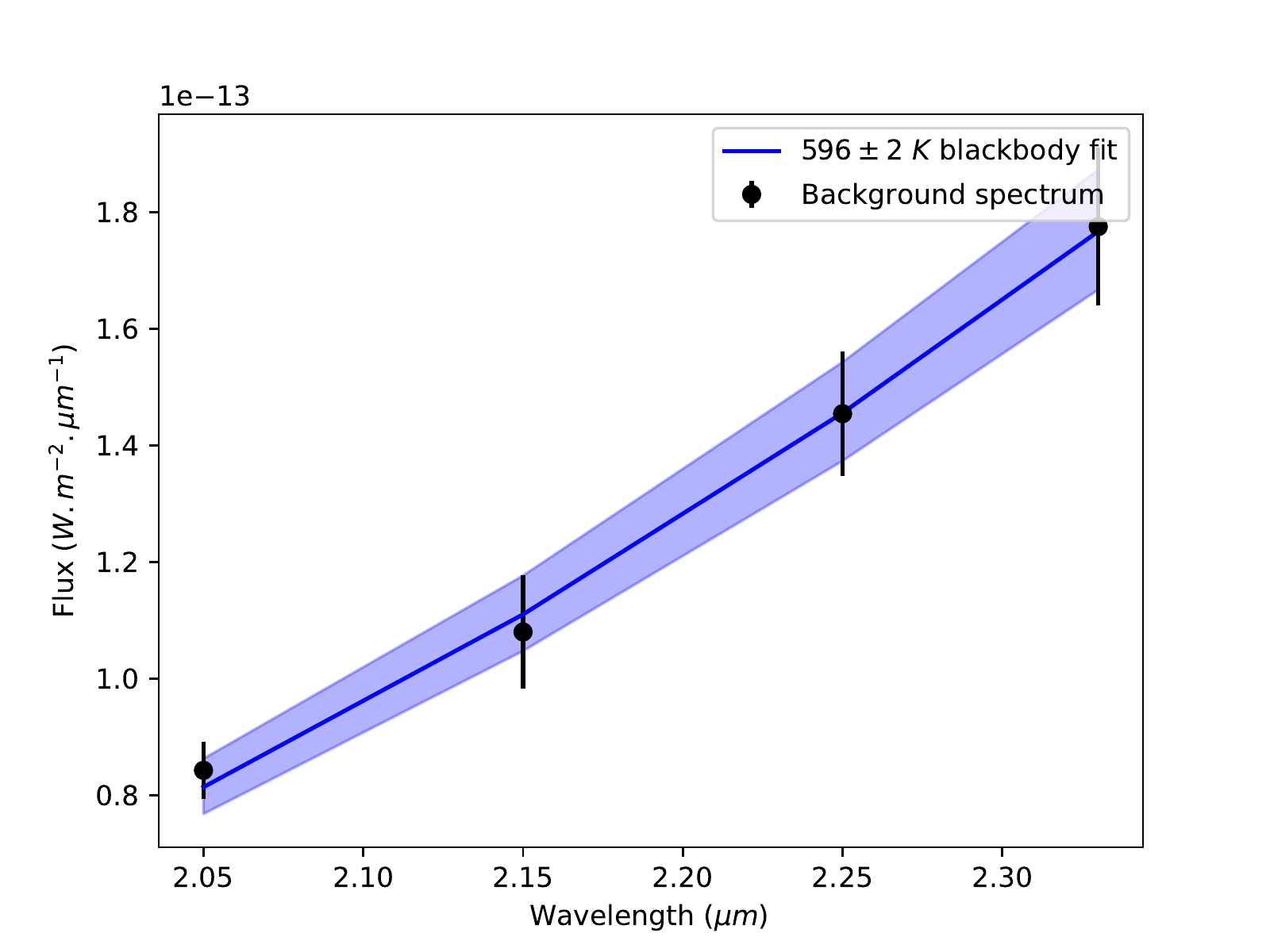}\label{fit_bg_2}}
	\caption{Spectral fits for model 2. \protect\subref{spec_cohe_2} : observed coherent spectrum and extincted model spectrum. \protect\subref{fit_bg_2} : Diffuse background and temperature estimation. }
	\label{specs2}
	\end{figure*}

	As can be noticed in Fig.~\ref{temp_mod2}, inside the dusty structure the temperature decreases very quickly with distance to the central source. This confirms that GRAVITY is observing solely the most internal region of the torus, the \textit{inner rim}. Figure~\ref{specs2} displays the spectra of the model for the compact source and for the diffuse background. The spectrum of the diffuse background is very well fitted by a $T \sim 600\ K$ black-body emission, which is consistent with the large scale structure temperature observed with MIDI in the mid-infrared (see Table \ref{comp}).
    
    \subsubsection{Unified model of the dataset}
    
    In the previous sections, we described several methods to interpret and model the GRAVITY data. Despite some diversity in the resulting parameters, recurrent features allow us to draw a synthetic description of the nuclear structures, that we present below before discussion and comparison with previous studies.
    
    First, the flux of the compact object represents around one-third of the total flux detected in the field of view of GRAVITY ($56\ mas$  or $3.9\ pc$), the remaining is considered as a diffuse component distributed on larger spatial scales.
    
    All the models agree on the characteristic size of this structure which is less than one parsec. More precisely, the different methods converge toward an elongated structure with a major axis of around $7\ mas$ ($0.5\ pc$) oriented northwest/southeast and with an elongation ration $e\sim3/2$.
    
    The apparent image of the source is an elongated ring. This is suggested by the simple geometrical models, the image reconstructions, and the various MontAGN models. This aspect can be explained in 3 dimensions by a roughly toroidal structure observed from a latitude comprised between $35\degree$ and $45\degree$.
    
    The best model found to describe the infrared emission is an optically thick structure with a geometrically thick disk shape and composed of graphite heated by a central UV-X source. The inner radius of this disk, which coincides with the sublimation region of graphite, is comprised between $0.20\ pc$ and $0.25\ pc$, which corresponds to a central source of luminosity comprised between $3.9\times 10^{38}\ W$ and $6.4\times10^{38}\ W$. As we assume that the dust is at its sublimation temperature and that the central radiation is fully responsible of its heating, this value should be taken as an upper-limit of the central UV-X luminosity. Since this upper-limit is relatively close to independent other estimations and lower-limits \citep[between a few $10^{37}$ and a few $10^{38}\ W$ according to][]{Pier1994, BlandHawthorn1997, Kishimoto1999, Gallimore2001}, we confirm a posteriori that the dust must be close to its sublimation temperature and mostly heated by the central radiation.
    
    The closer and further edges do not have the same luminosity, which can be explained by the self-absorption effect of secondary photons. It is possible to get from this information an estimation of the density of the dust, which in turns provides an estimation of the mass dust of the structure, $M_{dust}\sim 1.75\ M_{\odot}$.
    
    Importantly, the whole region is obscured by a large quantity of interstellar dust located between the object and the observer.
    
    Reproducing the entirety of the observables, the MontAGN model  2 is considered the \emph{best model} in the following and the parameters presented in Table \ref{model_2} will be used as reference for the following discussion unless otherwise specified.

\section{Discussion on the overall torus structure}

The orientation found in this study for the hot dust structure seems incompatible with results from previous observations of NGC 1068's nucleus. However, as will be discussed afterward, the geometry and the dynamic of the torus at parsec scale is much more complex than what can be suggested by lower angular resolution observations. This observational complexity reveals an asymmetry, and possibly high turbulence or instabilities in the heart of the torus. In this perspective, two interpretations will be discussed to explain the observed discrepancies: a unique unstable disk, and an apparent superposition of entangled rings.

\subsection{Discussion on the central structure inclination and orientation}

Every methods presented in this paper to model the GRAVITY data indicate an orientation of the structure comprised between $135\degree \lesssim  PA \lesssim 150\degree$ and an inclination $40\degree \lesssim i \lesssim 70\degree$. The best model has $PA=150_{-13}^{+8}$\,$\degree$ and $i=44_{-10}^{+10}$\,$\degree$.

Though, several observations, as described below, suggest either an obscured nature of the nucleus, associated with an inclination around $90\degree$ well in line with the classical unified model of AGN \citep{Antonucci1993}, or orientation of the central structure different from ours.

First of all, the historical spectro-polarimetric observation of NGC 1068 \citep{Antonucci1985} revealed a polarized Seyfert 1 emission in the heart of NGC 1068, interpreted as a type 1 nucleus hidden behind a large amount of dust \citep{Antonucci1993}. This structure is assumed to be roughly toroidal -as it is the case for our model- but seen close to edge-on for an obscured AGN such as NGC 1068. At first sight, our results seem to contradict this model.

At the few ten parsec scale, the \textit{extended torus} has been observed. Its outskirt, very dimly lit, was observed thanks to polarimetric techniques in \citet{Gratadour2015}. The image of the object is elongated, with dimensions $60\ pc \times 20\ pc$, which suggests an edge-on structure. This signature of the extended torus is oriented with $PA=118\degree$ which differs by $15\degree$ to $30\degree$ from our estimations.

At sub-parsec scales, the VLBA radio continuum observations \citep{Gallimore2004} also suggest the presence of an elongated structure, S1, which could trace an edge-on disk, with $PA \sim 110\degree$. Moreover, the detection of many maser spots located on a line crossing the nucleus indicates without any ambiguity the presence of an edge-on disk, with $i\sim90\degree$. Nevertheless, this maser disk is not oriented with $PA\sim110\degree$ like other structures mentioned so far, but with $PA\sim 135\degree$, which is compatible with the lowest estimations obtained from GRAVITY data. 

As we will see in the remaining of the discussion, the misalignment between these two radio structures (maser disk and S1) is also observed between structures at larger scales. To support this discussion, inclination measurements as well as position angle for the various observations are synthesized in Table \ref{comp}.

\paragraph{A structure with $PA\sim145\degree$}

As the previously mentioned maser orientation \citep{Greenhill1997}, several observations have revealed the presence of one or more elongated structures oriented along $PA\sim135\degree-140\degree$. In particular, the $800\ K$ dust observed with MIDI is oriented with $135-140\degree$, exactly as the $1600\ K$ hot dust observed with GRAVITY in this study. Also, \citet{LopezGonzaga2014} highlight a polar emission (located north of the central source), colder and more extended, with $PA\sim145\degree$. We can also note that the highly polarized ridge observed in \citet{Gratadour2015} is approximately orthogonal to this $PA\sim140\degree$ orientation, of which it could trace a polar counterpart.

\paragraph{Inclinations}

Only few observations  provide an actual measurement of the inclination of the torus at parsec and sub-parsec scales.
Indeed, $i\sim90\degree$ found in literature is more often an hypothesis motivated by the obscured nature of the nucleus or by the orientation of the NLR and jet rather than a direct measurement. As explained in \citet{Nixon2013}, the orientation of a jet originating from accretion around a black hole is very stable as it is linked to the rotation axis of the massive object. Its orientation is not expected to show any significant changes on timescales $\tau\leq10^7\ years$, even in the case of chaotic accretion. In comparison, the timescale of a cloud orbiting at $0.5\ pc$ from the central mass is $\tau \sim 2000\ years$. In consequence, there is no guarantee that the orientation of the dusty structures coincides with the equatorial plane perpendicular to the jet and NLR axis.

At first sight, the elongated shapes observed with ALMA or MIDI also suggest structures with low inclination. However, the observed major and minor axes ratios, $2 \lesssim a/b \lesssim 3$ \citep{Gallimore2004,LopezGonzaga2014,Gratadour2015} do not really exclude inclinations close to $60\degree$.

Beside the $i=44_{-10}^{+10}$\,$\degree$ estimation obtained from the 3D MontAGN model presented in this study, only \citet{GarciaBurillo2016} from ALMA data and  \citet{Greenhill1997} provide an estimate at those spatial scales. The first one finds an estimation of the inclination by fitting a CLUMPY torus model \citep{Nenkova2008} on the observed SED. When limiting the inclination to values comprised between $60\degree$ and $90\degree$, the authors found $i=66\degree$, which could be compatible with GRAVITY results on hot dust. But when removing the a priori constraint on the inclination, the best fit becomes $i=33\degree$, which is even less inclined than our models. This estimation is anyhow in much better agreement with the GRAVITY results than with an edge-on structure.

At last, the most challenging difference is between our measurement of the inclination and the presence of the maser disk. Indeed, the detection of masers is conditioned by the presence of a coherent flux of gas directed toward the observer. It gives information on the inclination, which has to be close to $i=90\degree$ \citep[inclined by at most $2\degree$ from this configuration according to][]{Gallimore2001}. Also, since a maser spot is observed as a point source, its position is known with the astrometric resolution of the VLBA, which is of high quality. Hence, both $i$ and $PA$ are known with precision for the maser disk. Moreover, these precise measurements allowed to characterize the maser disk with more details and to conclude it has a $0.65\ pc$ inner radius, larger than the outer radius of the hot dust structure (see Figs. \ref{im_source_mod2} and \ref{temp_mod2}).  Thereby, these two structures do not necessarily occupy the same volume of space and their simultaneous existence is possible. 

\paragraph{Summary}

The inner region of the torus appears complex, with at least three favored planes, which exhibit close but significantly different orientations :
\begin{enumerate}
\item The extended molecular torus plane ($\sim 10\ pc$), with $PA\sim110\degree-120\degree$ and a inclination which is not measured precisely but presumed close to edge-on ($i\sim90\degree$).
\item The maser disk plane, with $  PA\sim135\degree$ and $i=90\degree$. The $800\ K$ dusty structure observed with MIDI could be in that plane.
\item the hot dust structure observed with GRAVITY, with $PA=150_{-13}^{+8}$\,$\degree$ and $i=44_{-10}^{+10}$\,$\degree$.
\end{enumerate}

Overimposed to this complex structuration, several dynamical structures are observed: a high turbulence as well as non-circular motions \citep{GarciaBurillo2016, Imanishi2018}, the outflow which interacts with the ISM as close as $0.6\ pc$ from the nucleus \citep{Gallimore2016}, and the recently observed velocity field that suggest a counter-rotation between the maser and molecular disks \citep{Imanishi2018, Impellizzeri2019}. We note that the latter observation could also be explained by an outflowing torus model \citep{Garcia-Burillo2019}.

Far from the simple picture of a unique structure in equilibrium or a regular inflow toward the central mass, the torus appears to be a turbulent region, where cohabit and survive several structures with very different geometries and dynamics.

\subsection{A unique unstable disk or several intricated rings}

The observed diversity of sizes and orientations could be explained by at least two models: either the various structures observed with different orientations are actually belonging to a single object, with a warped disk shape, or are indeed different entangled structures, with roughly circular shapes.

\paragraph{Warped disk}

If the inner region of the torus observed at the parsec scale constitutes the prolongation of the accretion disk, it likely has the shape of a disk at bigger spatial scales presenting asymmetries and distortions to explain the different observations. Two instabilities are known to be able to affect tori and accretion disks.

The first one, called \textit{runaway instability} to highlight its cataclysmic effects \citep{Abramowicz1983,Abramowicz1998}, is caused by an axisymmetric perturbation. It happens when the accretion disk overflows in the Roche lobe of the central mass, i.e., the black hole mass as well as the mass of surrounding gas, which is significant \citep{Lodato2003}. The inflow of matter produced toward the central mass pushes the Roche lobe further away, leading to an exponential increase of the mass transfer and the accretion of the whole structure in a few dynamical times (for NGC 1068, a few thousand years for parsec scale structures) \citep{Korobkin2013}. 

The second instability known to possibly affect AGN happens when a disk is submitted to low order non-axisymmetric perturbations (for example a significant mass present on one side of the disk). Called PPI instability, it was proposed by \citet{Papaloizou1984} and vastly studied since then. It leads to a transfer of the angular momentum from the inside to the outside of the disk, producing density asymmetries \citep{Bugli2018}. This instability can lead to a \textit{runaway} scenario.

The apparition of instabilities in a disk may explain the variety of orientation which is observed in the heart of NGC 1068, while keeping a unique structure to describe the object.

However, there is no published study on the maser spots that indicate the presence of a warped disk, as it is the case for other objects \citep[see][for the warped disk of Circinus for instance]{Greenhill2003}, and in addition it appears unlikely that either a \textit{runaway} or a PPI instability could produce counter-rotating outer and inner disks as actually observed. The very quick propagation and development of these instabilities until reaching the critical moment when the accretion material is depleted (a few dynamical timescales) makes the observation of the phenomenon unlikely, and is difficult to conciliate with the continuous activity of the AGN highlighted by the jet and NLR sizes. The stability of such a structure could only be sustained by the presence of a binary system of supermassive black hole \citep{Wang2020}.

\paragraph{Entangled rings}

A second interpretation of the incompatible parsec scale observations  actually relies on different structures, originating from different clouds orbiting the central mass, and having different orientations, radial distances, motions, and chemical compositions.

One of the most plausible scenario to explain the feeding of an AGN is the continuous collision of clouds constituting the torus that lose their angular momentum, fall into the gravity well and provide material for the accretion \citep{Sanders1981}. Once a cloud goes crosses the Roche limit, it is torn apart by the tidal forces and can form a disk or ring structure. Similarly, \citet{Impellizzeri2019} suggest that the capture of a molecular cloud or a dwarf satellite could explain the presence of a counter-rotating disk. Depending on the density of clouds and their velocity dispersion, the number of collisions and the life expectancy of these structures can greatly vary. At the $10\ pc$ scale, two of these \textit{tongues} of matter are detected flowing to the nucleus from the northern region \citep{Sanchez2009}.

Some of these rings formed by the disruption of clouds could survive up to sub-parsec distances and be separately detected by the different mentioned instruments  (SPHERE, ALMA, VLBA, MIDI, GRAVITY). 

This model offers a lot of freedom for interpretation and can explain easily the various orientations observed. Moreover, it may explain the counter-rotation observed between the maser disk and the molecular torus, that would have arisen from two clouds with distinct orbits \cite{Imanishi2020}.

\paragraph{Conclusion}

Both models provide an explanation for the variety of observed orientations. However, the unique warped disk scenario cannot explain the presence of an inner counter-rotating region and faces difficulties to maintain accretion on long time scales. The second model can explain the various observations, including the counter-rotations. The formation of the rings is realistic, even if their survival at these small spatial scales is surprising. Globally, the multiple entangled rings model is favored.

\section{Conclusions}

The GRAVITY observation on which is based this paper offers for the first time the possibility to study hot dust at the smallest spatial scale of the torus of NGC 1068. We show that a model based on realistic radiative transfer simulations provides a fair description of the observables. Most of the emission comes from a hot dust structure, which:
\begin{itemize}
\item Has an annular shape, with a $r=0.21\pm0.03\ pc$ radius. This ring appears to be geometrically thick, with an half-opening angle $\alpha_ {1/2}=21\pm8\degree$.
\item Is inclined with $i=44\pm10\degree$ and consequently does not obscure the central UV-X source. This result is surprising with regard of both the obscured nature of the central UV-X source and the observation of an edge-on disk at slightly large scales. 
\item Is aligned along $PA=150_{-13}^{+8}$\,$\degree$, which is consistent with the previous observation at parsec scale
\item Is overall obscured by a foreground extinction, leading to $A_V \geq 60$ ($\leftrightarrow A_K \geq 7$). 
\item Is constituted of graphite with a high density of grains ($n=10\ grains.cm^{-3}$ or $n=5_{-2.5}^{+5}\ M_{\odot}.pc^{-3}$), for a total of approximately one solar mass of dust.
\item Is dense enough to be optically thick to K band photons, explaining the contrast between the northeast and southwest edges.
\end{itemize}

We highlighted inconsistencies between its 3D orientation in space and the one from previously observed structures, which are also partially incompatible with each others. Two models could explain most of the differences: a unique warped disk, or a system of entangled rings. Since the warping of the disk is not observed in the astrometry of the maser spots and cannot explain the counter rotation, we favor the second model, where several rings have been formed from the tidal disruption of individual clouds.

\begin{acknowledgements}

      We thank the anonymous referee for its constructive comments. 
      
      This work was made possible by the doctoral school ED 127 \textit{Astronomie et Astrophysique d’Ile de France} which funded and accompanied with care the first author during its PhD. The last steps of the publication process were done while he was starting a position at the Astronomical Institute of the Czech Academy of Sciences, where he was supported by Czech Science Foundation Grant 19-15480S and by the project RVO:67985815.

      The Gravity data analyzed in this paper have been obtained thanks to ESO Large Programs IDs 0102.B-0667, 0102.C-0205 and 0102.C-0211. We would like to thank the Gravity collaboration, and more particularly our colleagues from Max Planck Institute for extraterrestrial Physics, for the fruitful discussions that allowed to improve our data analysis.
      
       We are grateful for the services provided by the NASA/IPAC Extragalactic Database, the SIMBAD database, and the various Python libraries used for this work, which include notably NumPy, SciPy, and AstroPy.
       
      \end{acknowledgements}
      

%
%

\bibliographystyle{aa} 
\bibliography{biblio} 




	

\begin{appendix} 
    \section{MontAGN models}
	
	\begin{figure*}[!ht]
	\centering		
	\includegraphics[width=0.49\hsize]{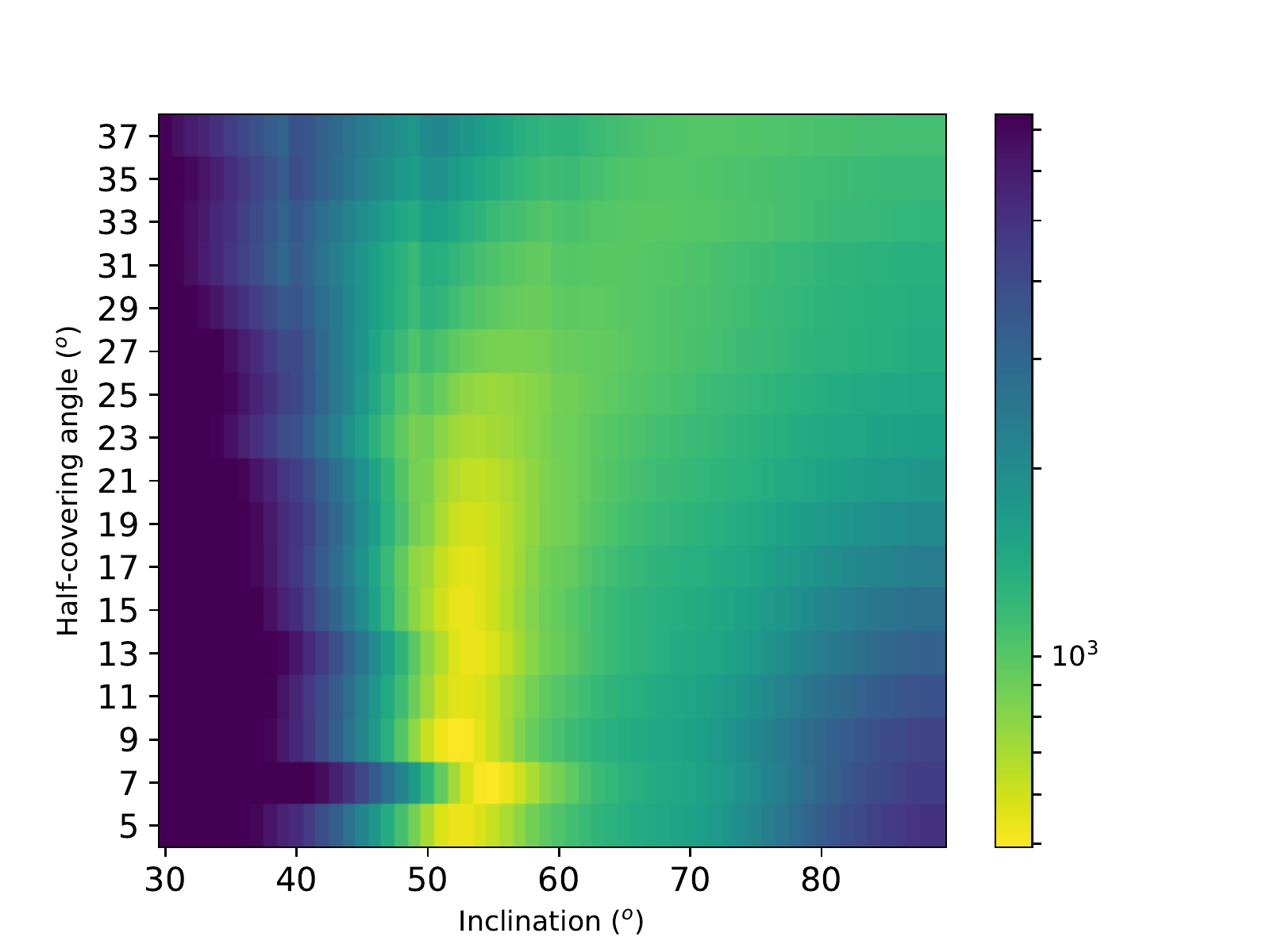}	
	\includegraphics[width=0.49\hsize]{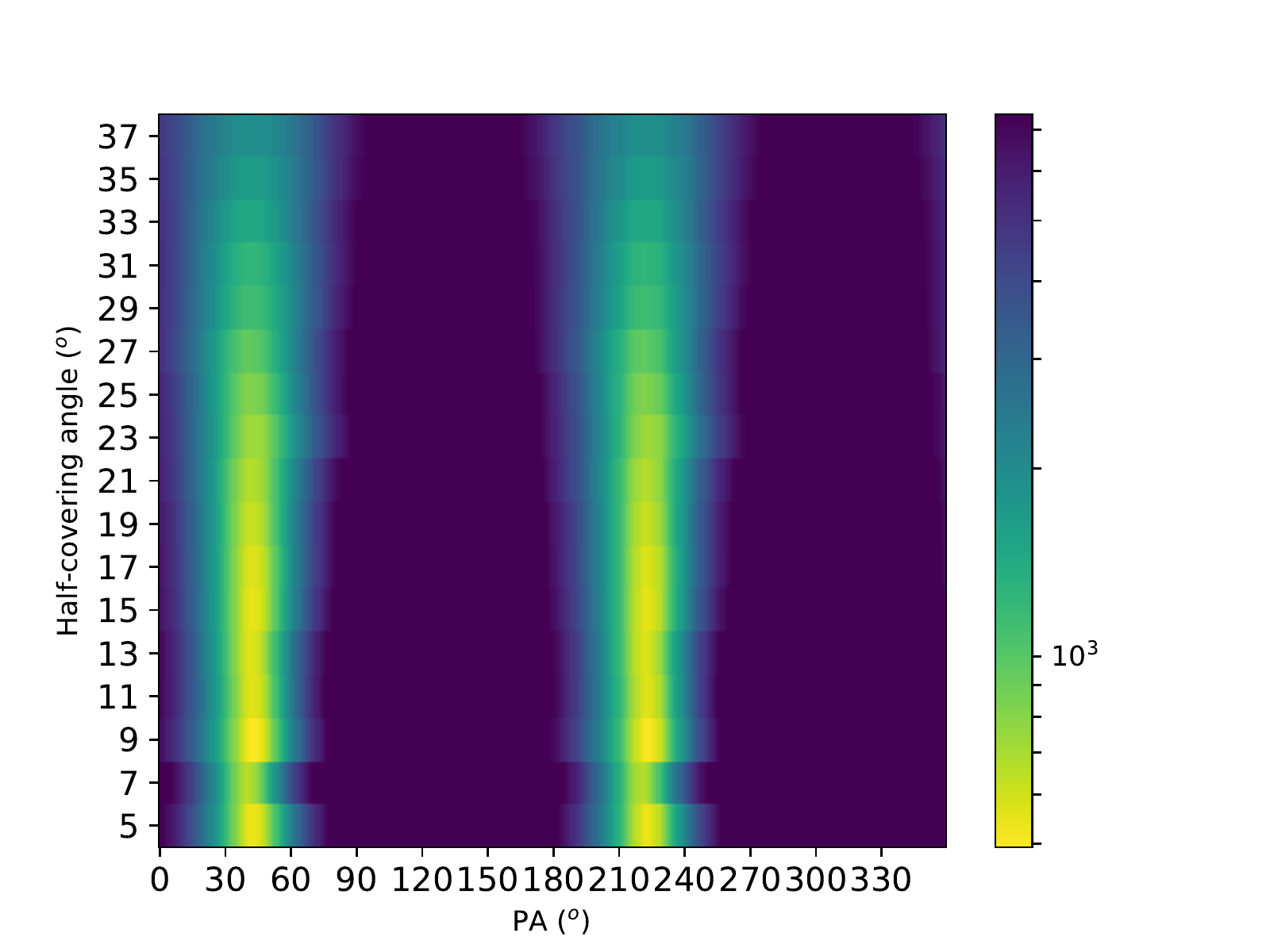}	
	\includegraphics[width=0.49\hsize]{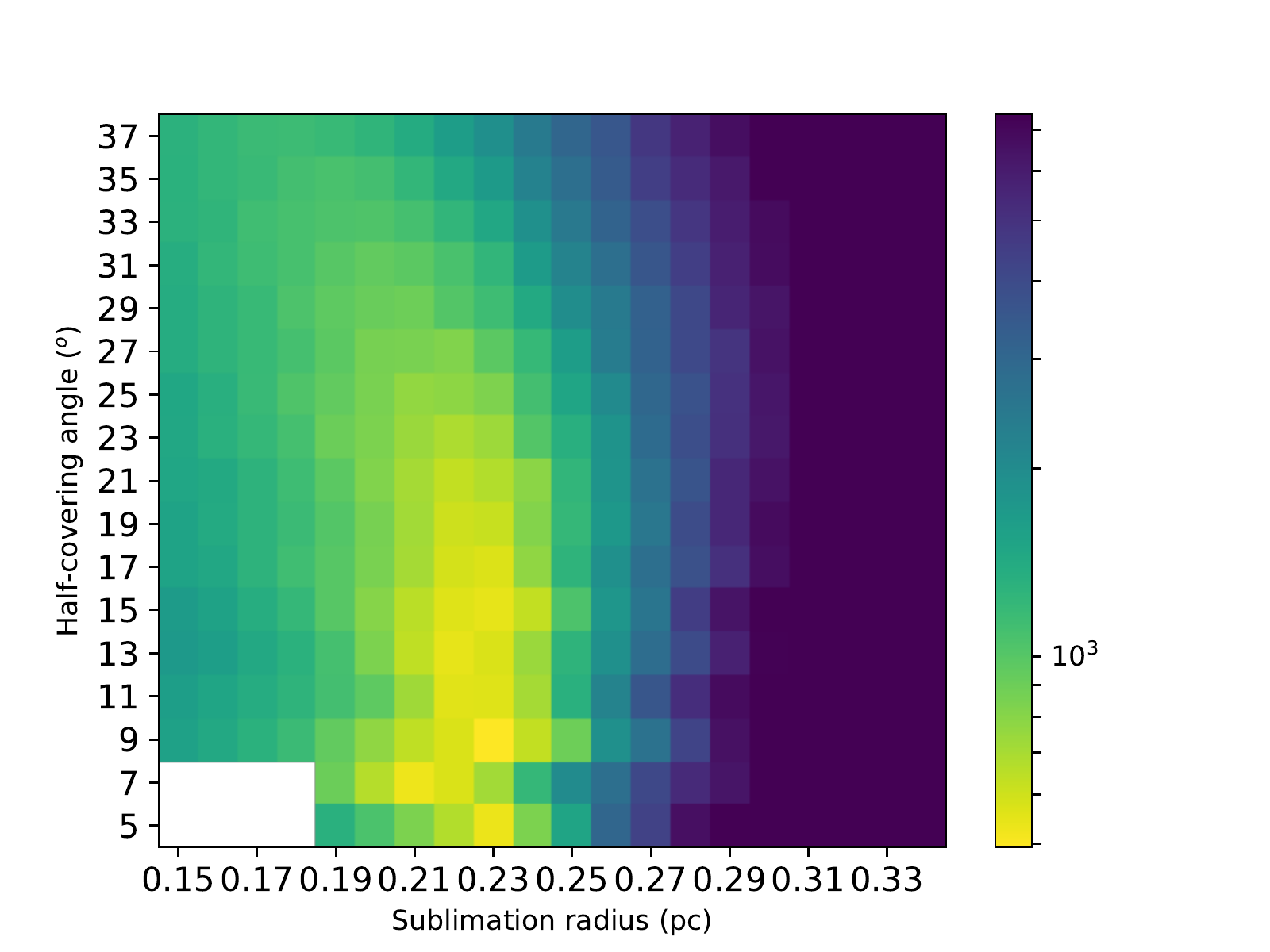}	
	\includegraphics[width=0.49\hsize]{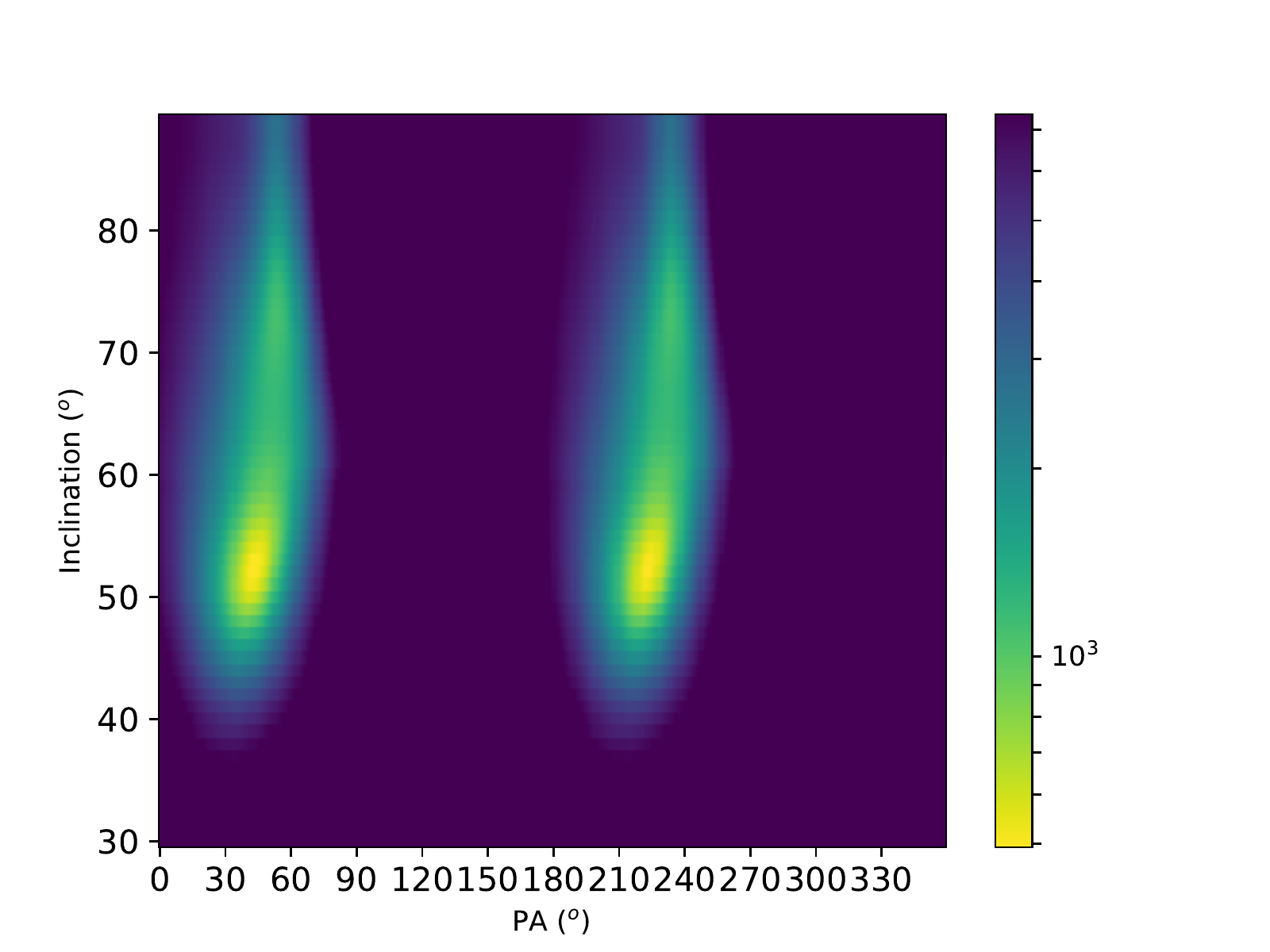}	
	\includegraphics[width=0.49\hsize]{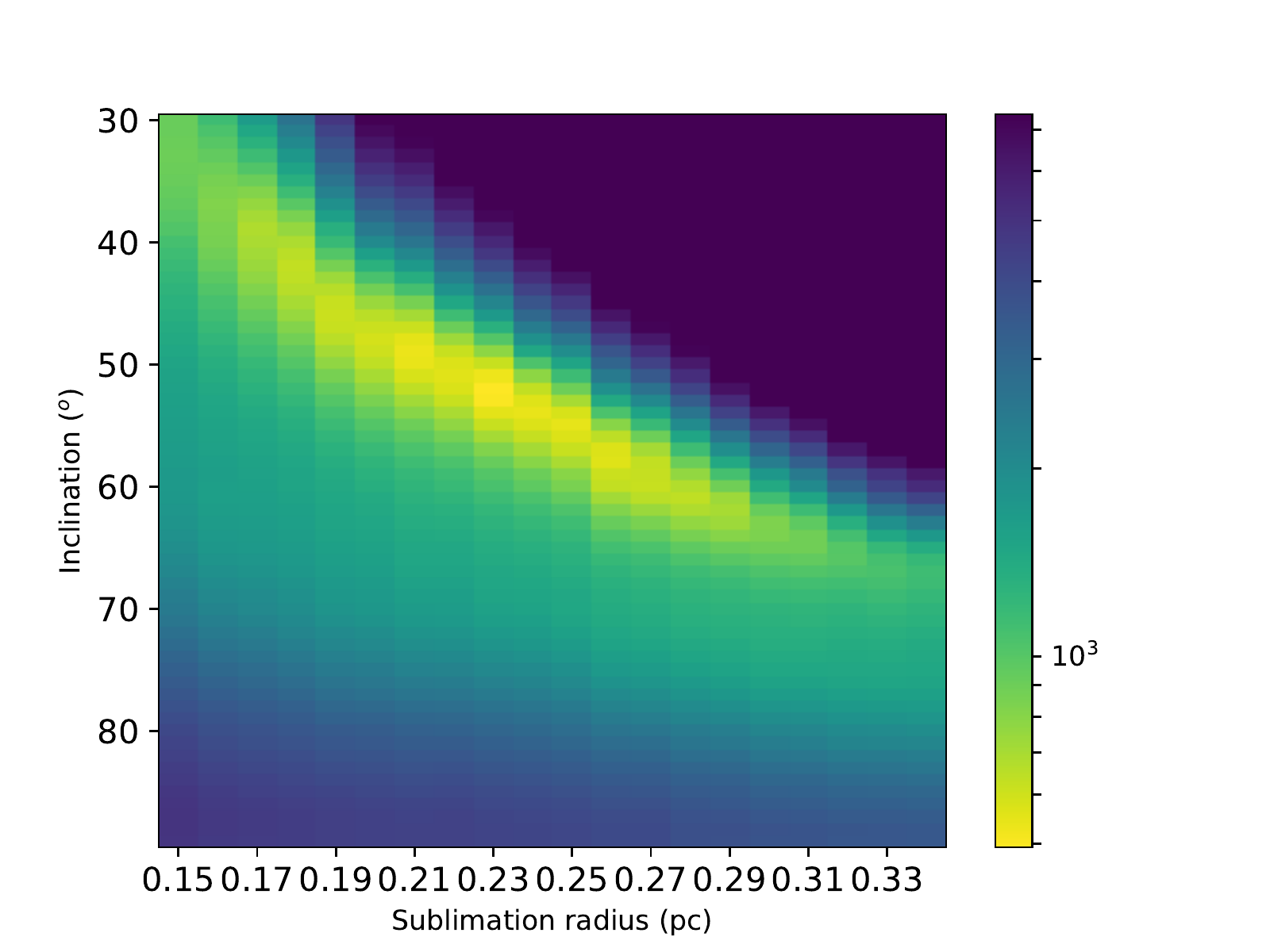}	
	\includegraphics[width=0.49\hsize]{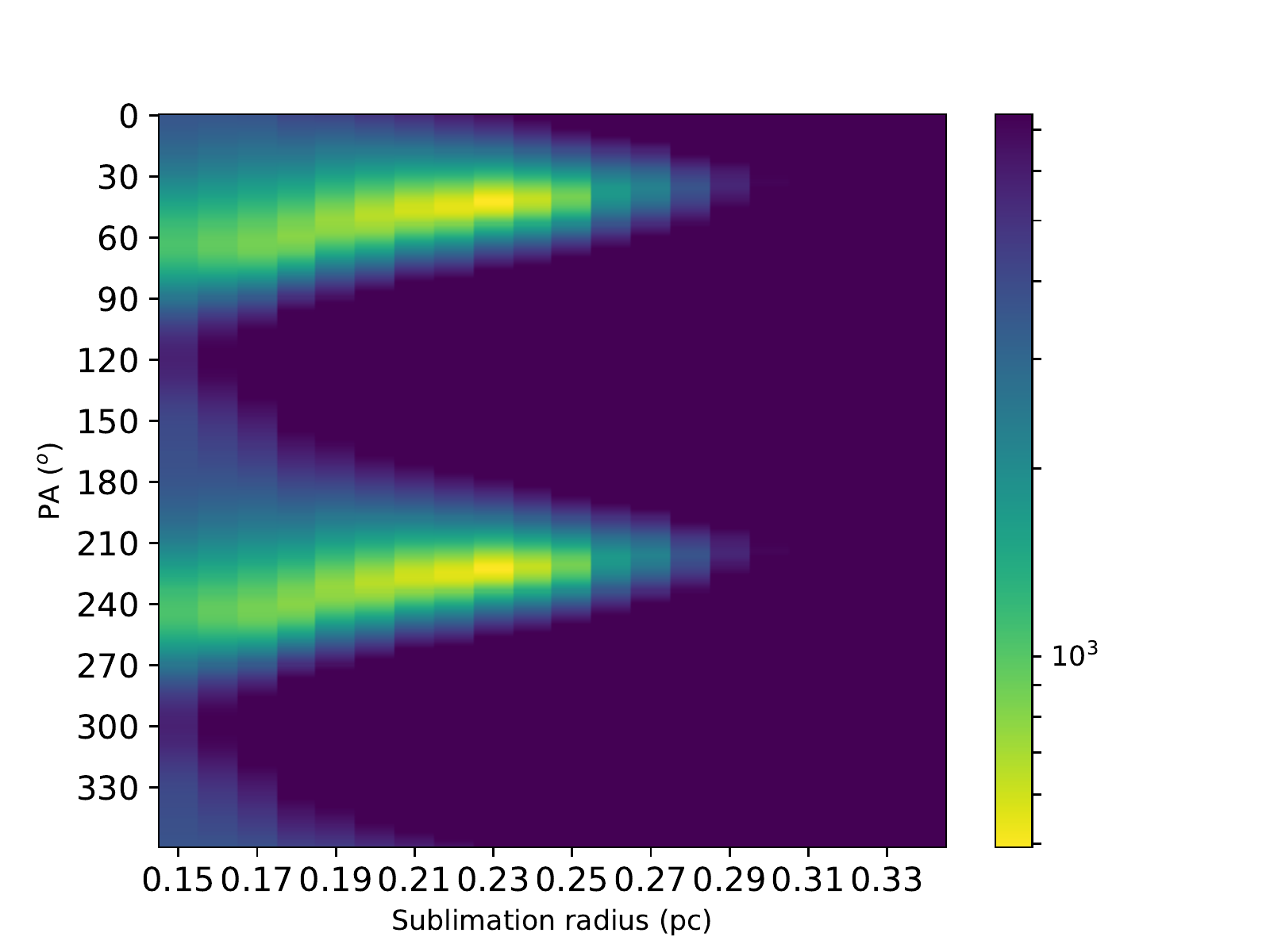}
	\caption{Cuts around the best solution in the $\chi^2$ cube obtained from MontAGN model  1}
	\label{chi2_1}
	\end{figure*}

	\begin{figure*}[!ht]
	\centering		
	\includegraphics[width=1.\hsize]{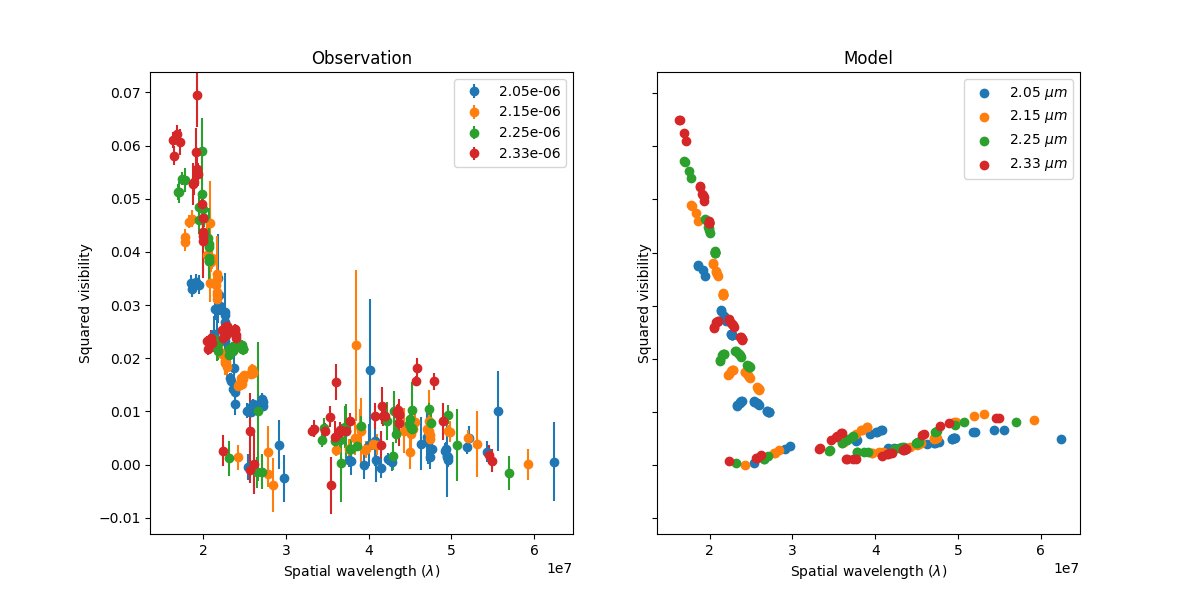}
	\caption{Left: Observed visibility, Right : Visibility from MontAGN model 1.}
	\label{vis_mod_1}
	\end{figure*}
	
	\begin{figure*}[!ht]
	\centering		
	\includegraphics[width=0.7\hsize]{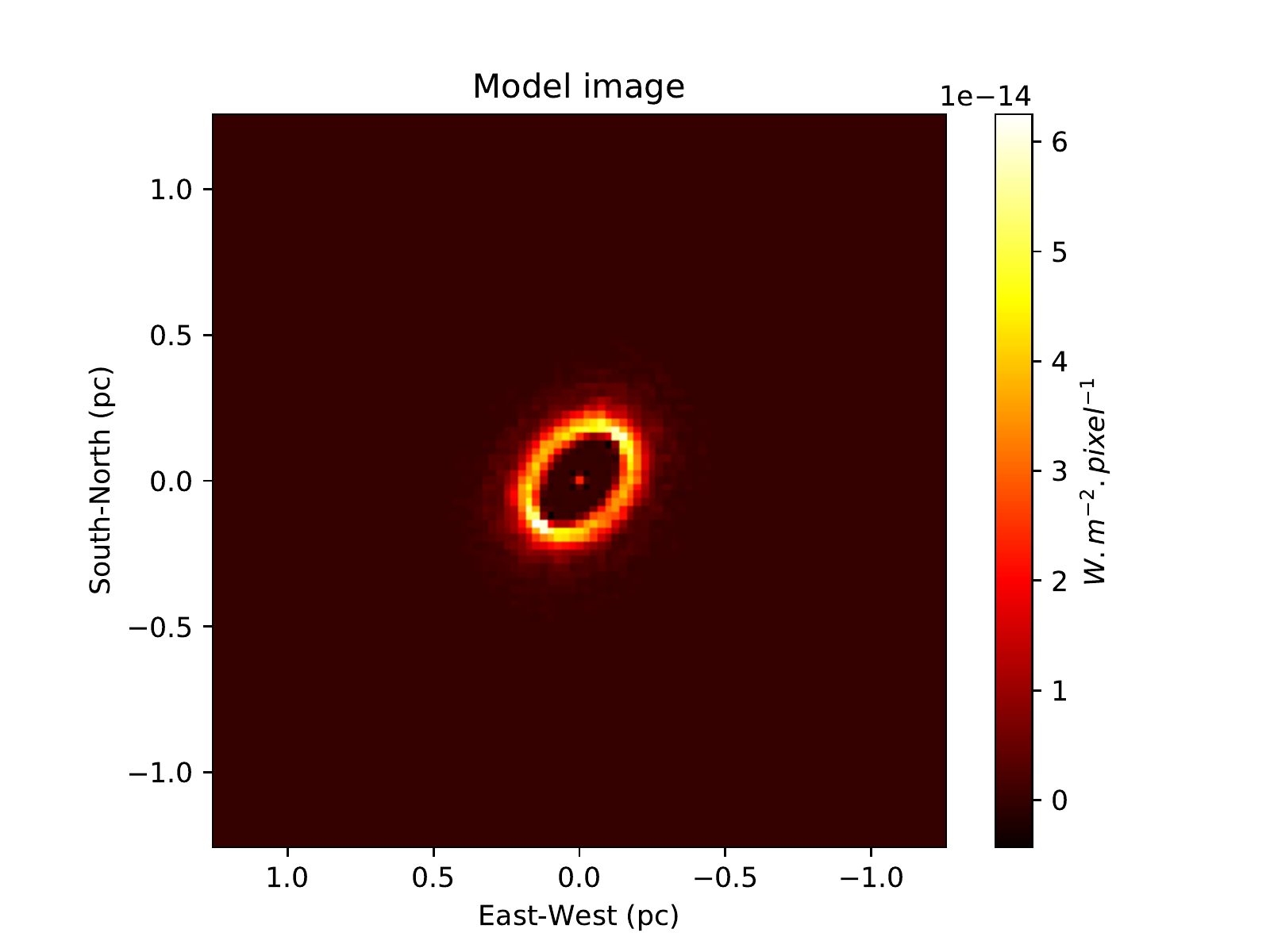}
	\caption{K-band image of the MontAGN model 1. The diffuse background is not represented.}
	\label{im_source_mod1}
	\end{figure*}
	
	\begin{figure*}[!ht]
	\centering		
	\subfloat[][]{\includegraphics[width=0.49\hsize]{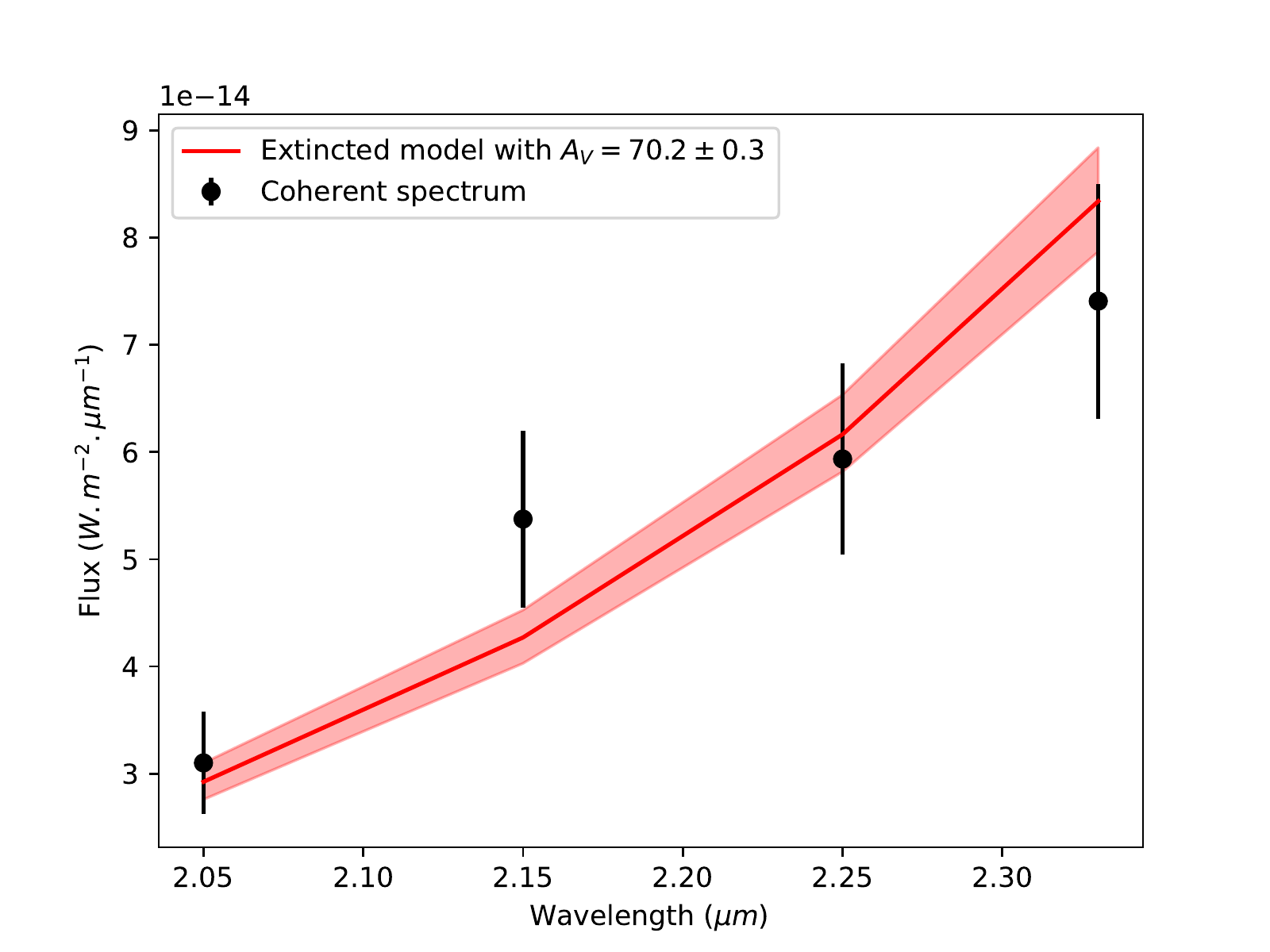}\label{spec_cohe_1}}
	\subfloat[][]{\includegraphics[width=0.49\hsize]{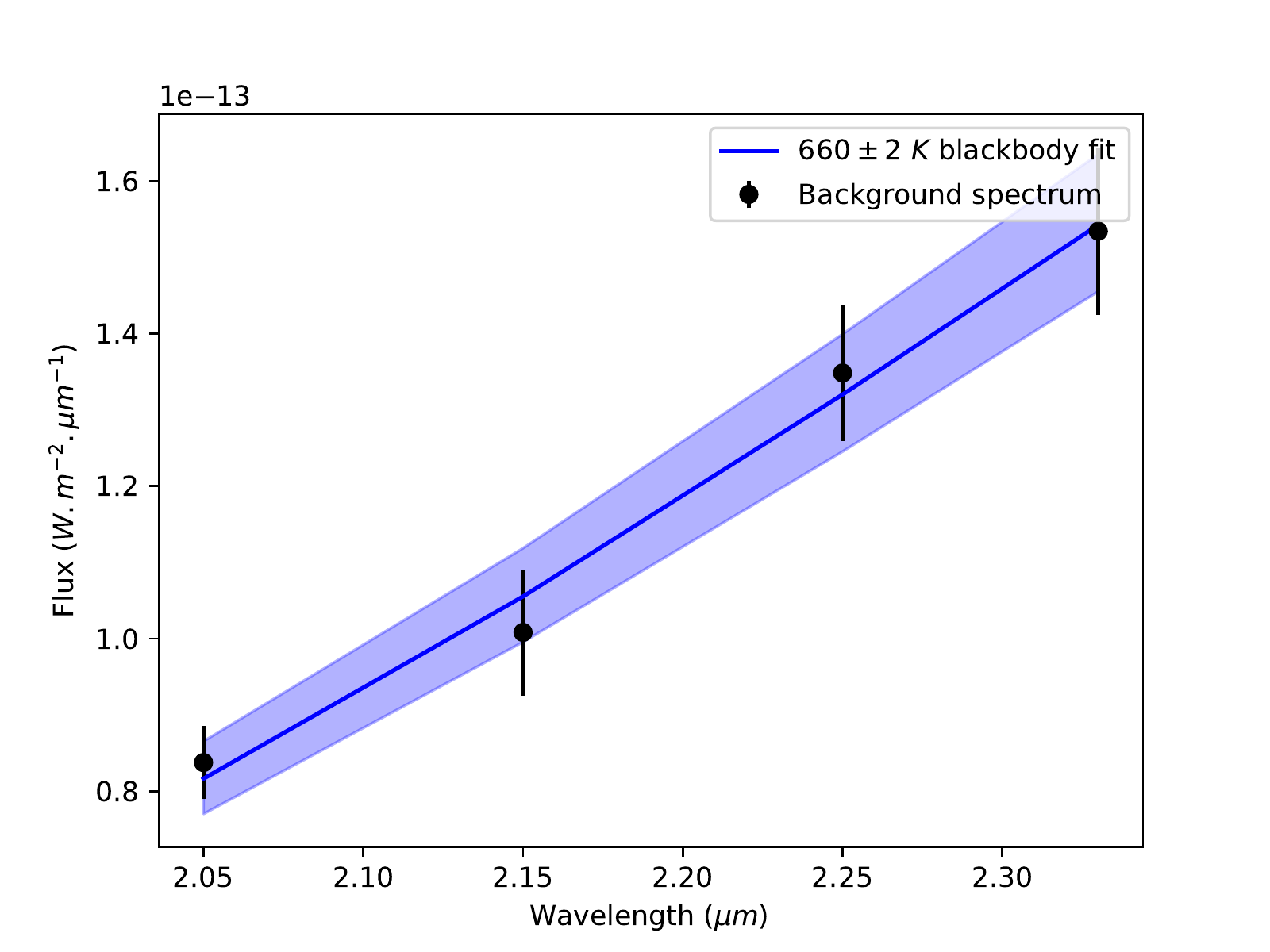}\label{fit_bg_1}}
	\caption{Spectral fits for MontAGN model 1. \protect\subref{spec_cohe_1} : observed coherent spectrum and extincted model spectrum. \protect\subref{fit_bg_1} : Diffuse background and temperature estimation. }
	\label{specs1}
	\end{figure*}
	
	\begin{figure*}[!ht]
	\centering		
	\includegraphics[width=0.49\hsize]{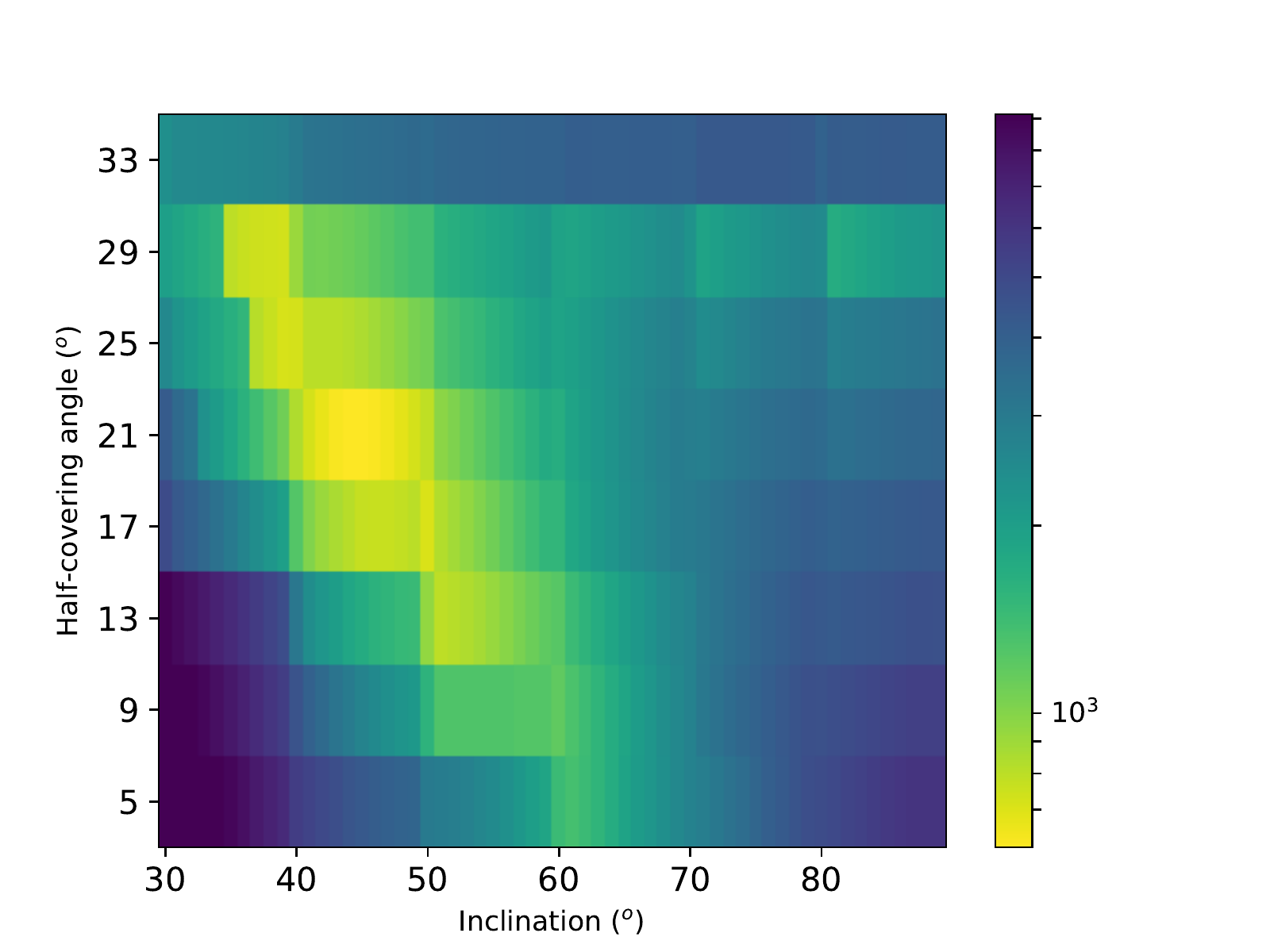}	
	\includegraphics[width=0.49\hsize]{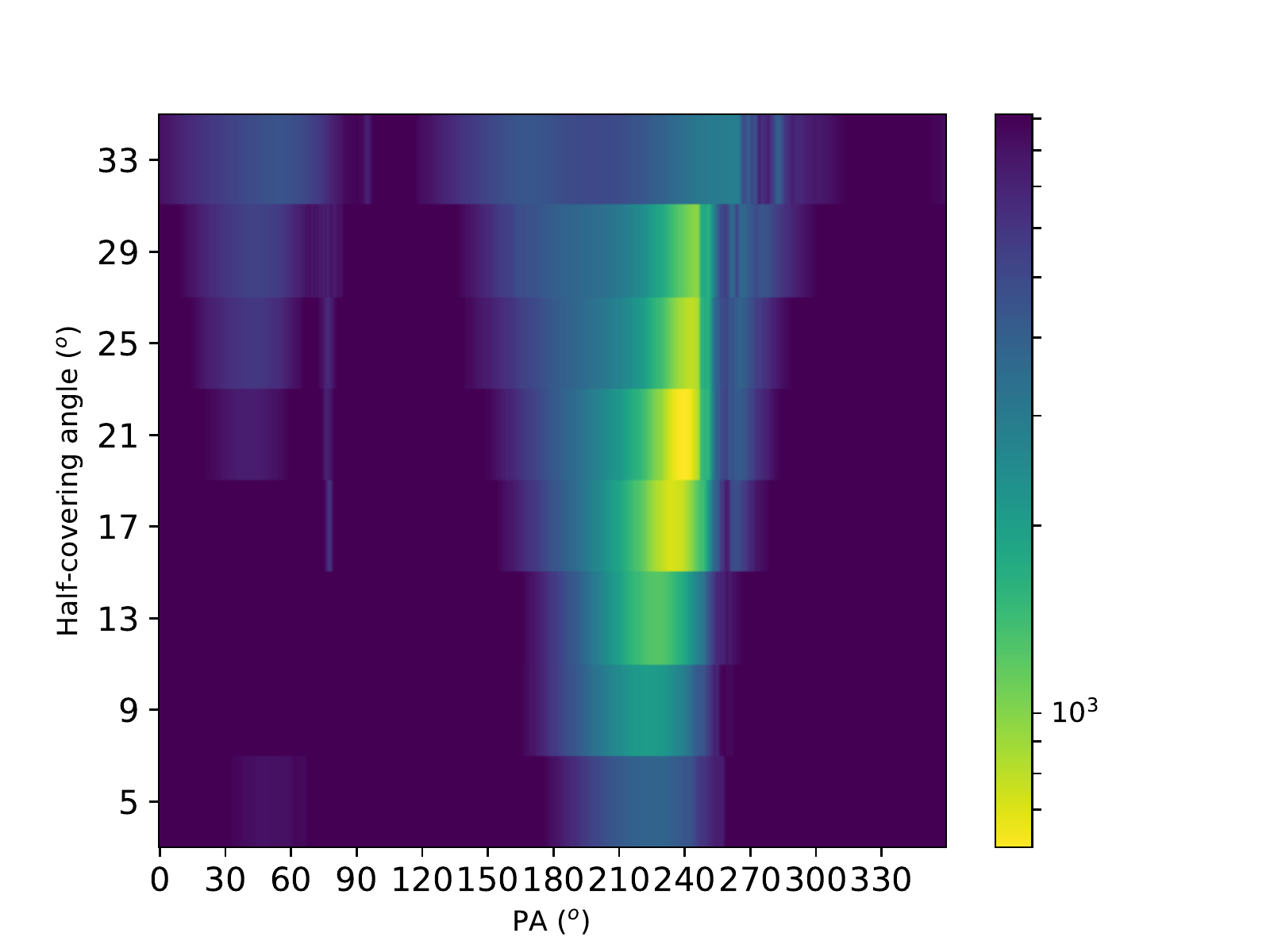}
	\includegraphics[width=0.49\hsize]{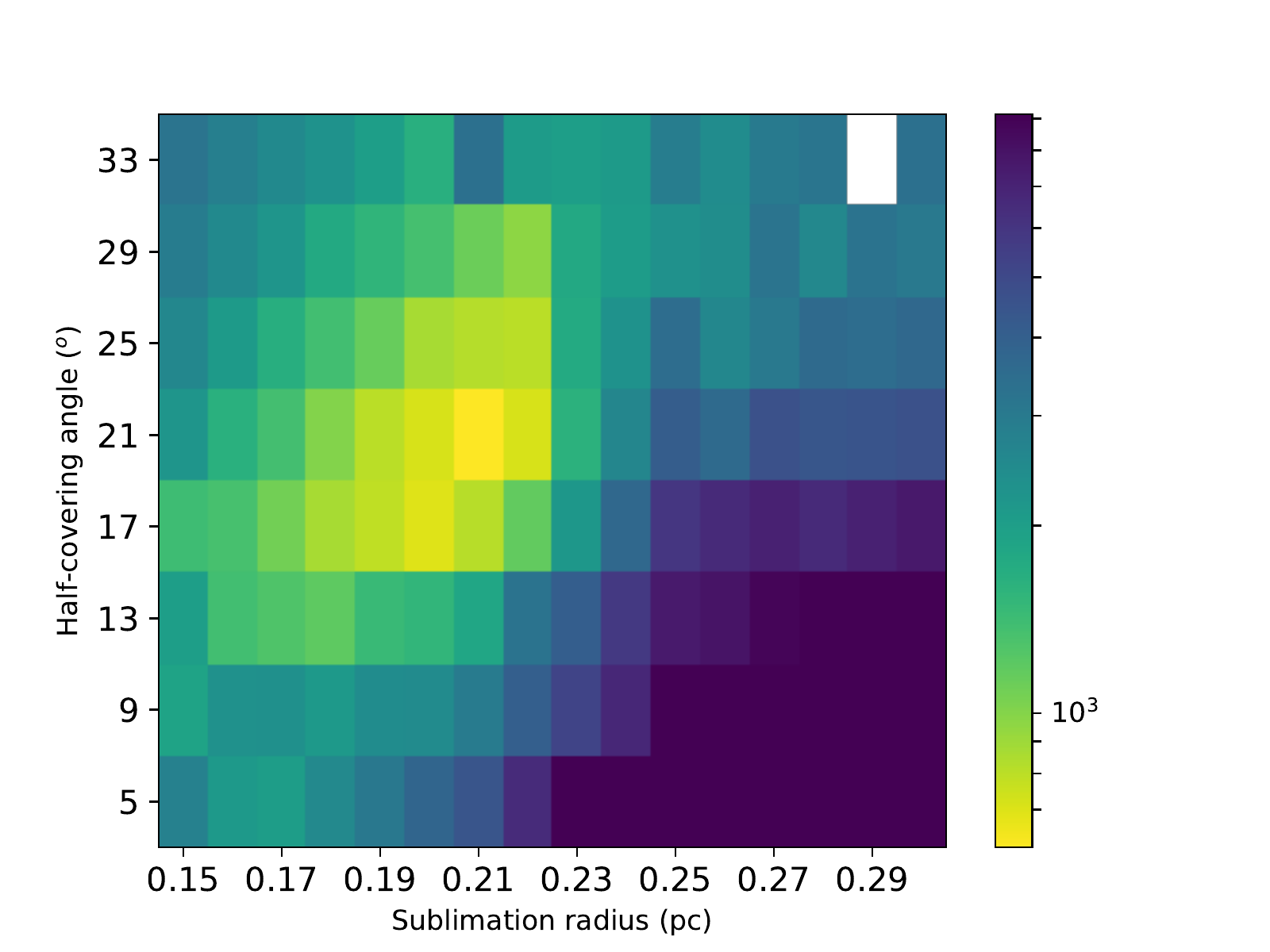}	
	\includegraphics[width=0.49\hsize]{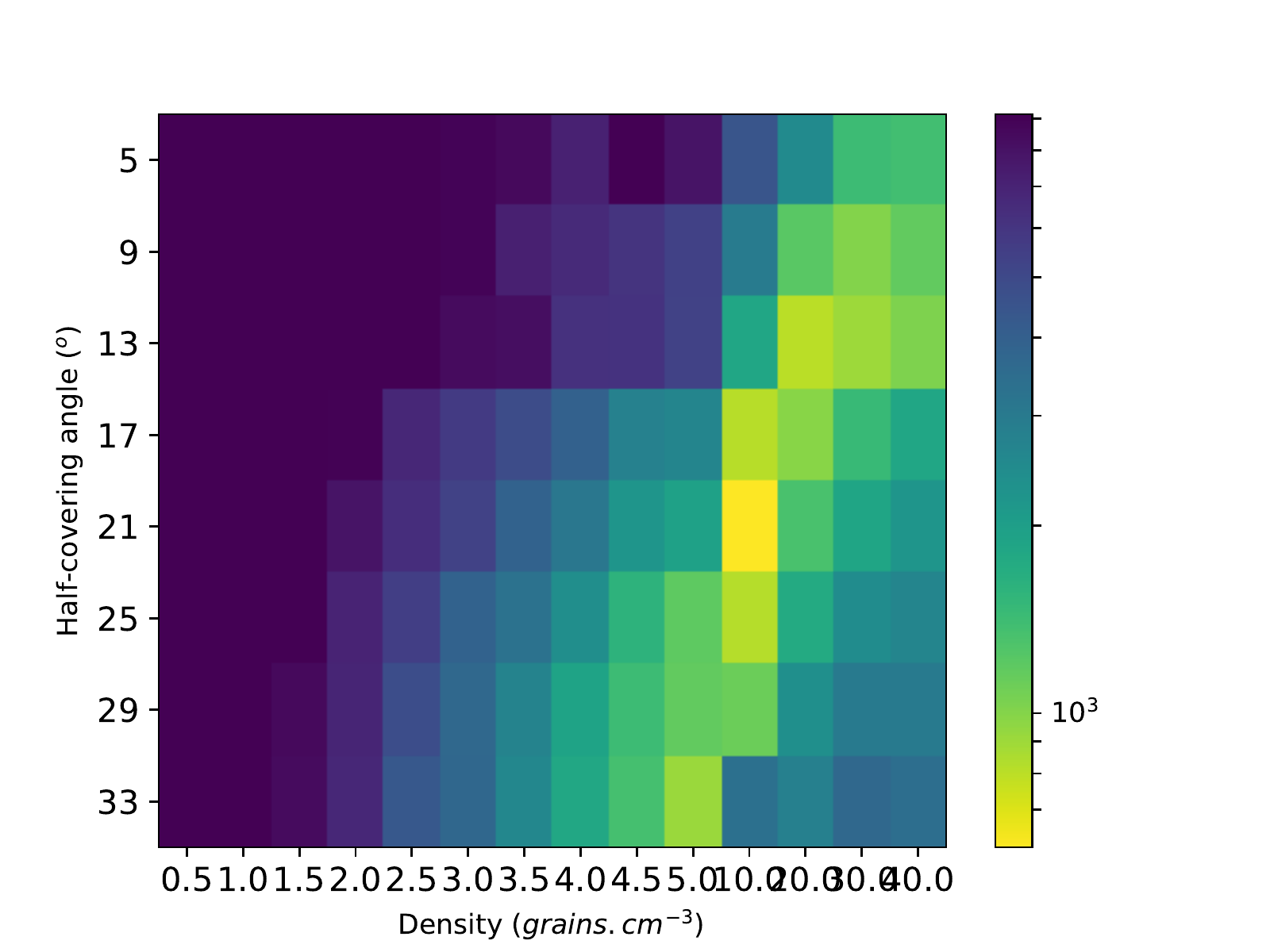}	
	\includegraphics[width=0.49\hsize]{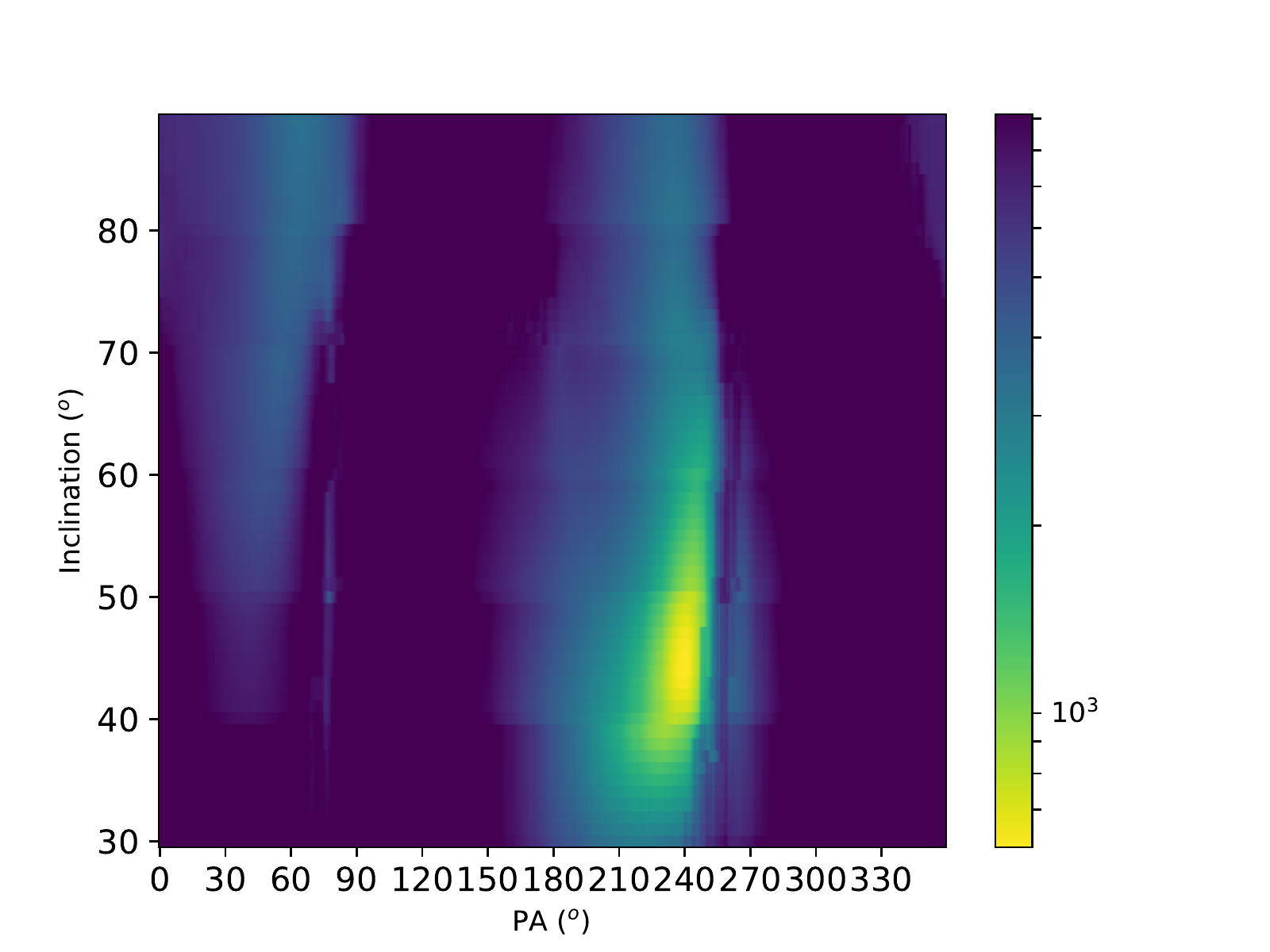}	
	\caption{Cuts around the best solution in the $\chi^2$ cube obtained from MontAGN model 2. Note that the dust is not sampled linearly.}
	\label{chi2_2a}
	\end{figure*}

	\begin{figure*}[!ht]
	\centering		
	\includegraphics[width=0.49\hsize]{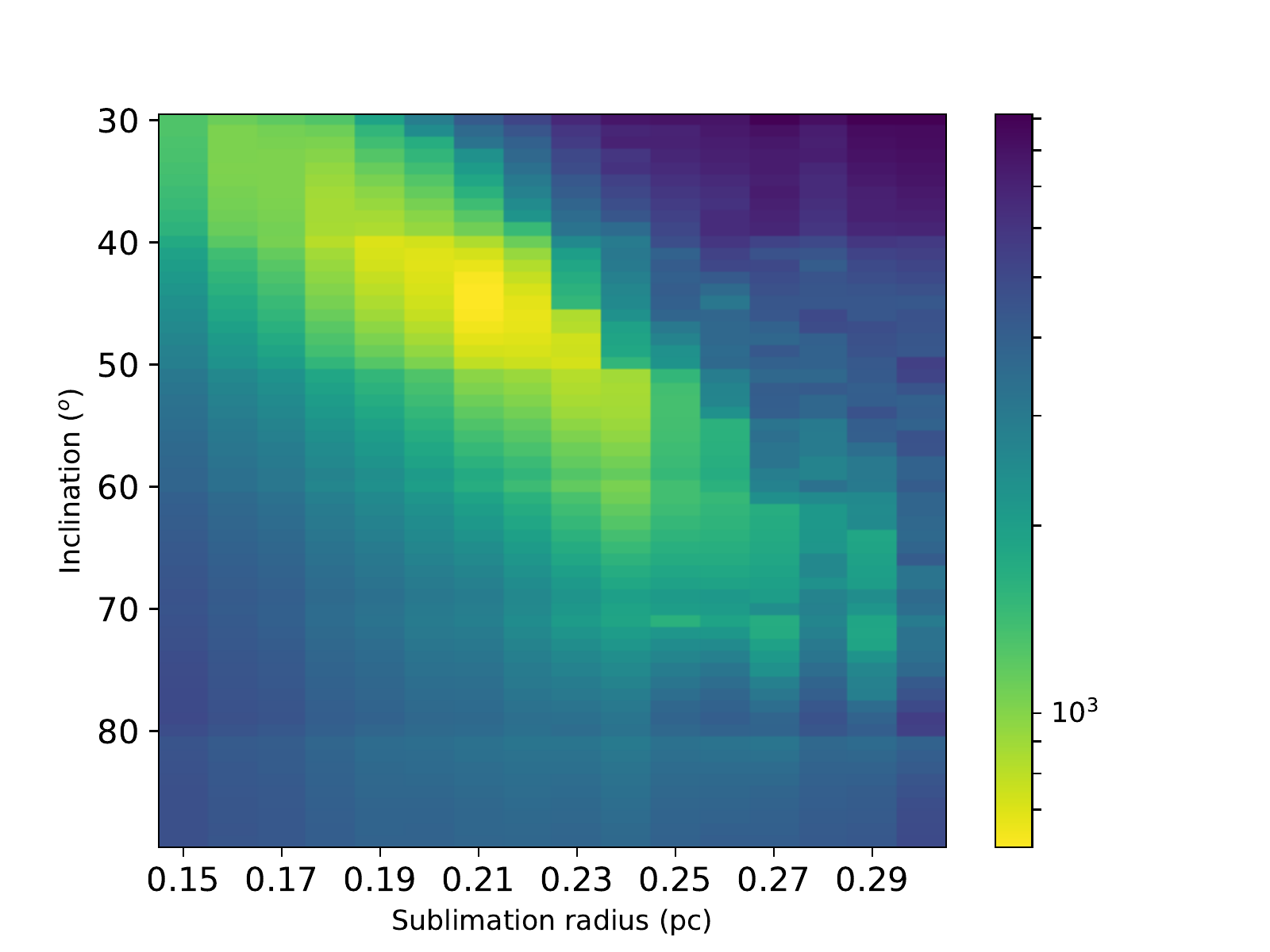}
	\includegraphics[width=0.49\hsize]{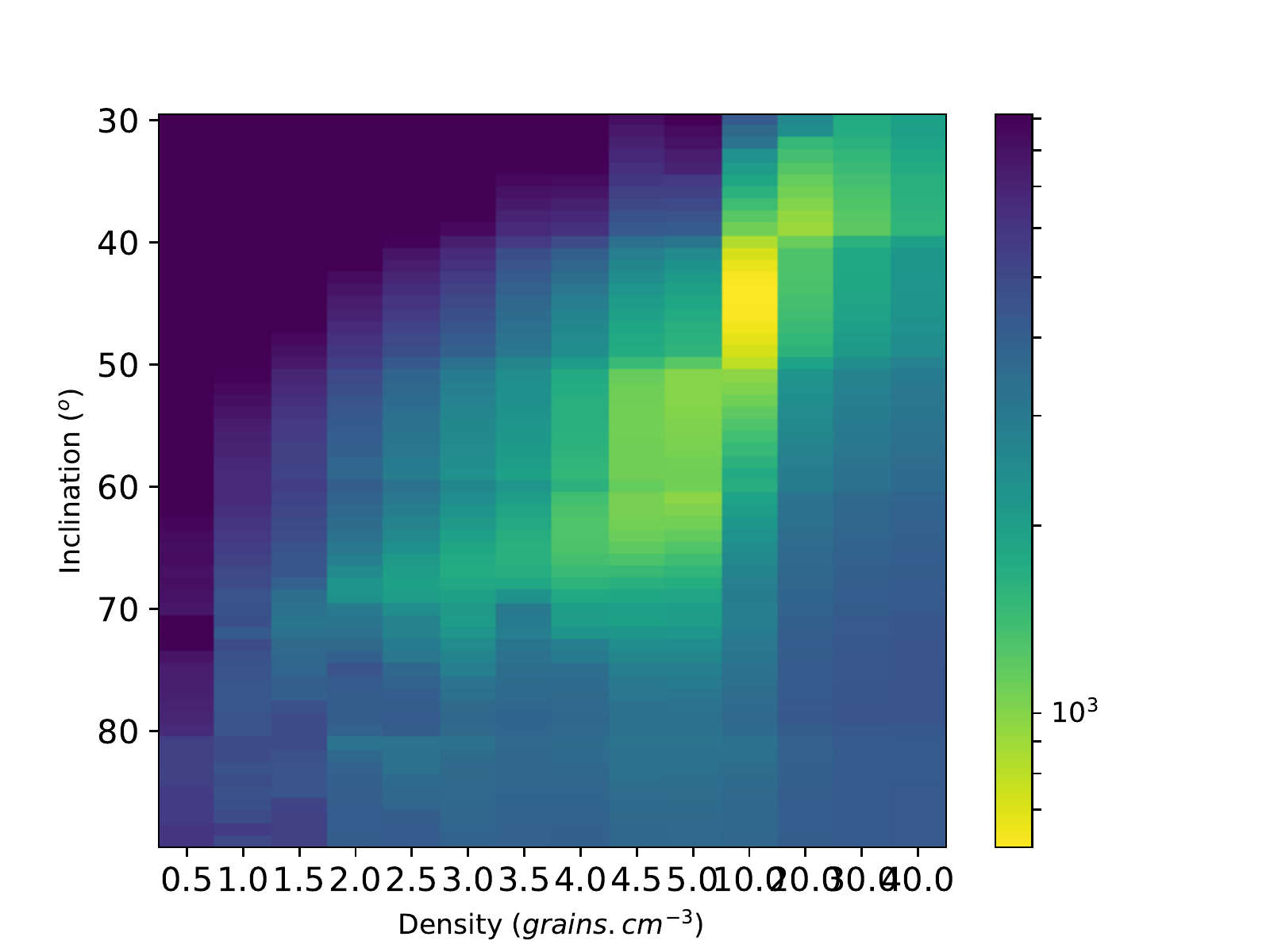}	
	\includegraphics[width=0.49\hsize]{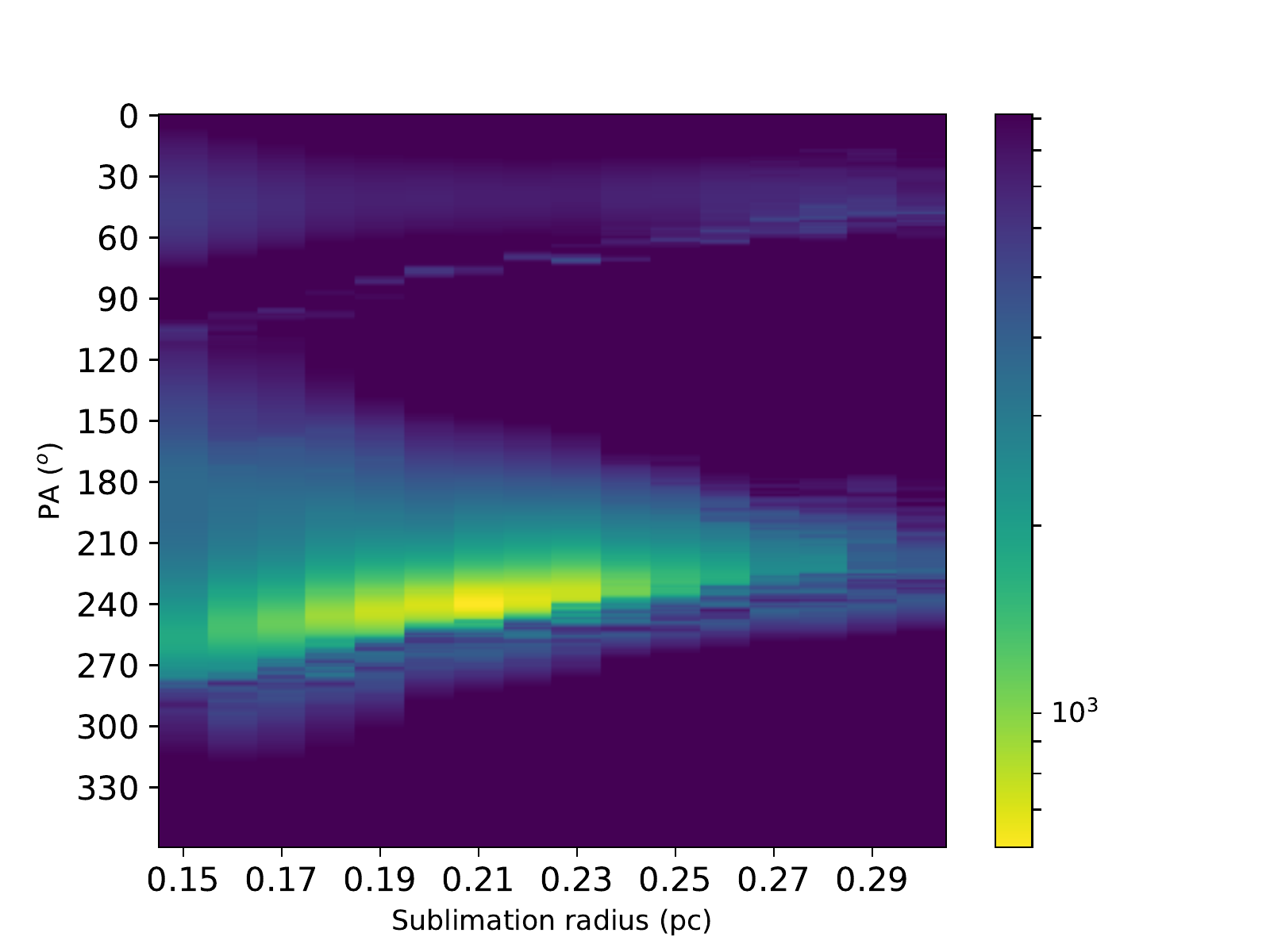}
	\includegraphics[width=0.49\hsize]{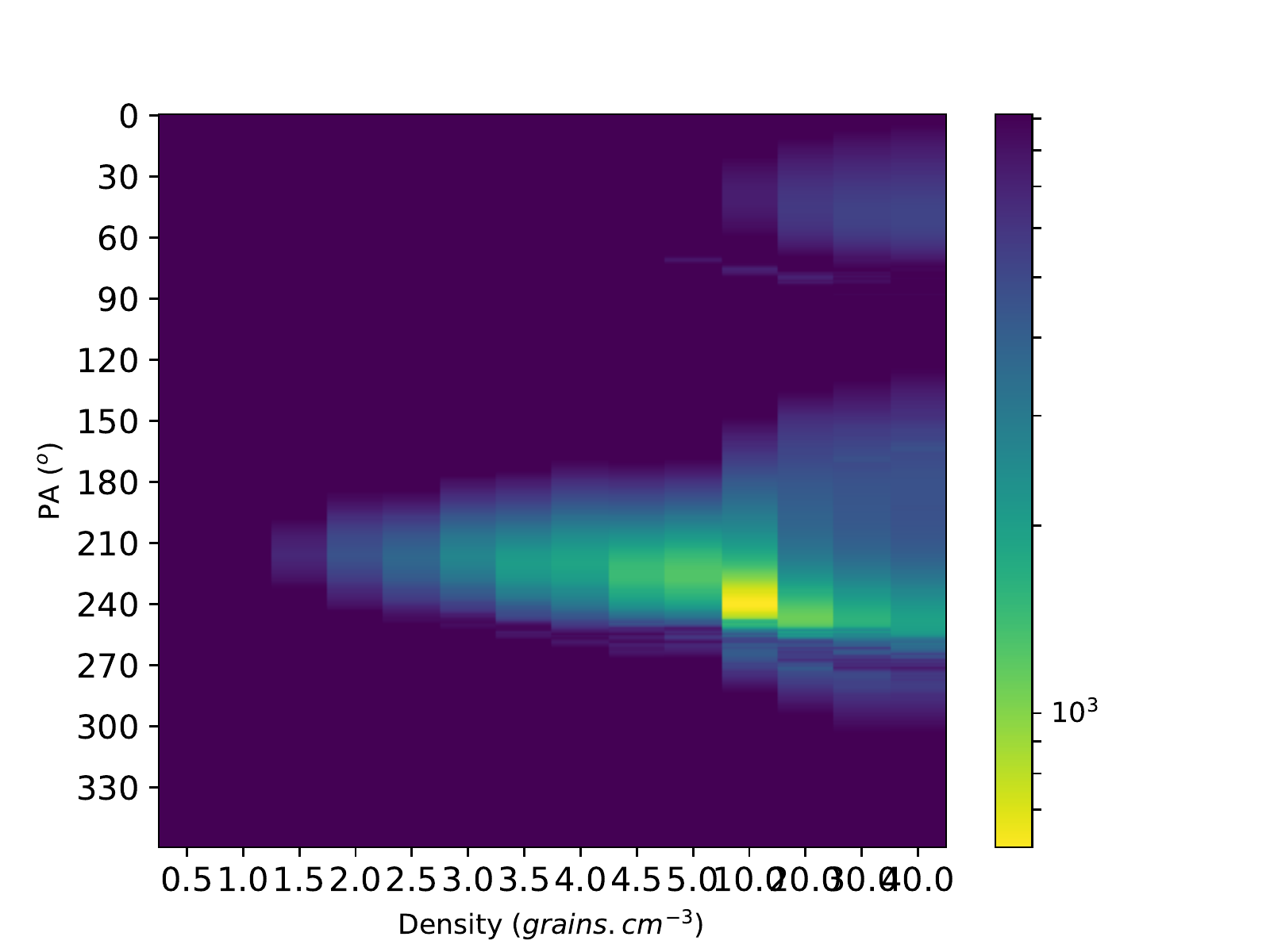}
	\includegraphics[width=0.49\hsize]{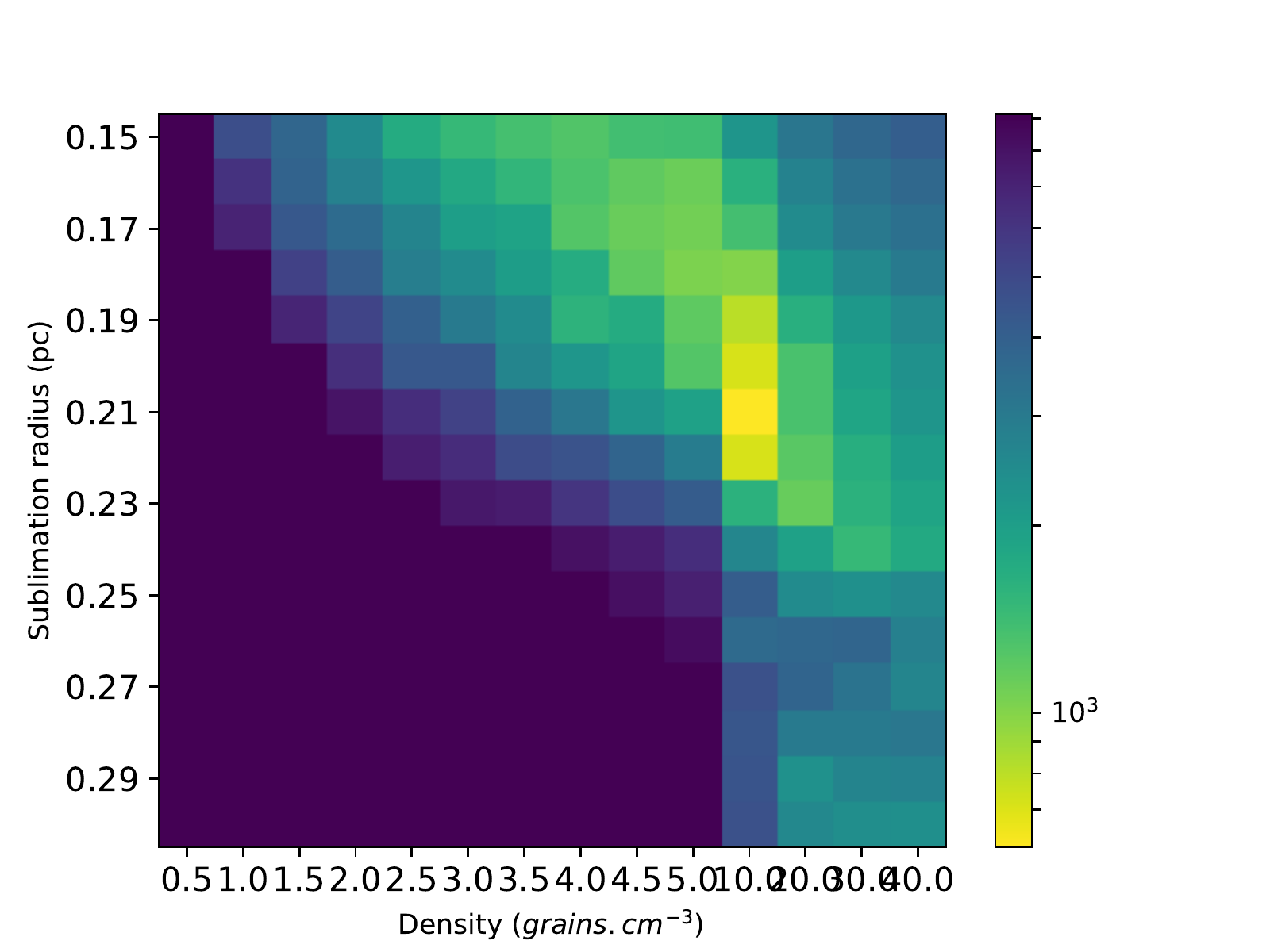}
	\caption{Cuts around the best solution in the $\chi^2$ cube obtained from MontAGN model 2. Note that the dust is not sampled linearly.}
	\label{chi2_2b}
	\end{figure*}

	\begin{figure*}[!ht]
	\centering		
	\includegraphics[width=1.\hsize]{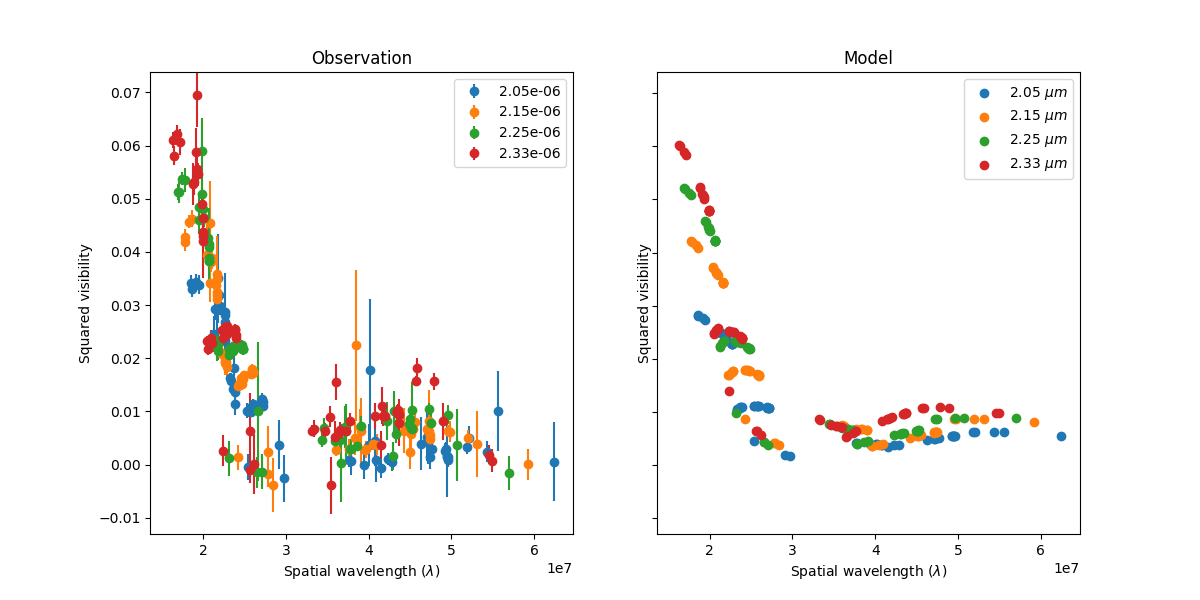}
	\caption{Left: Observed visibility, Right : Visibility from model 2.}
	\label{an_vis_mod_2}
	\end{figure*}
	
	\begin{figure*}[!ht]
	\centering		
	\includegraphics[width=1.\hsize]{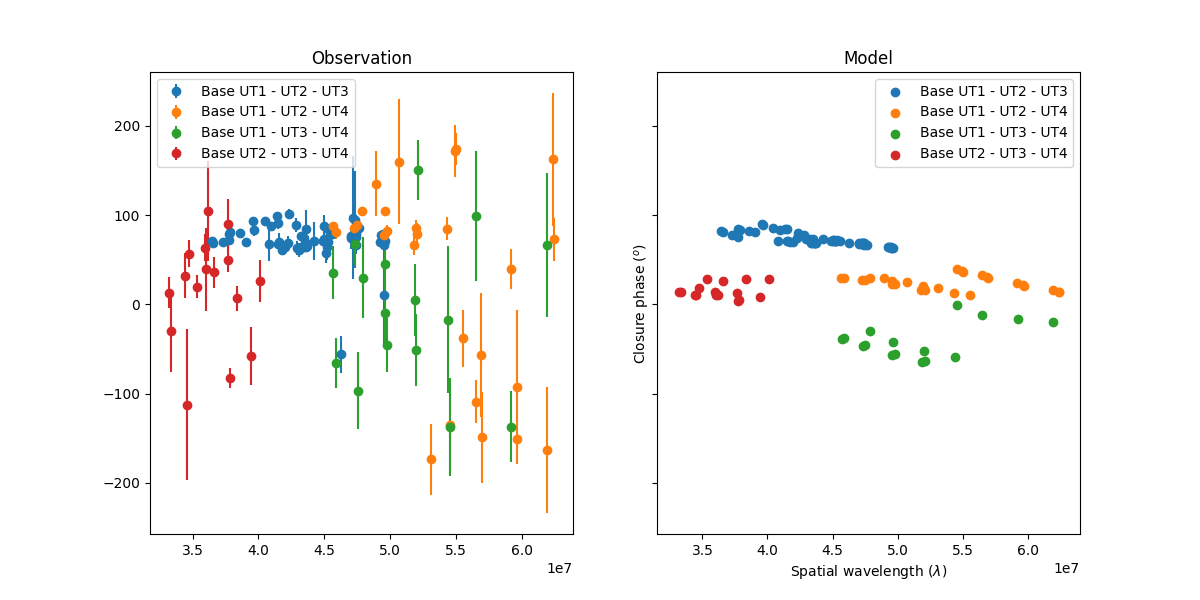}
	\caption{Left: Observed closure phase, Right : Closure phase from MontAGN model 2.}
	\label{an_t3_mod_2}
	\end{figure*}

	\begin{figure*}[!ht]
	\centering		
	\includegraphics[width=0.7\hsize]{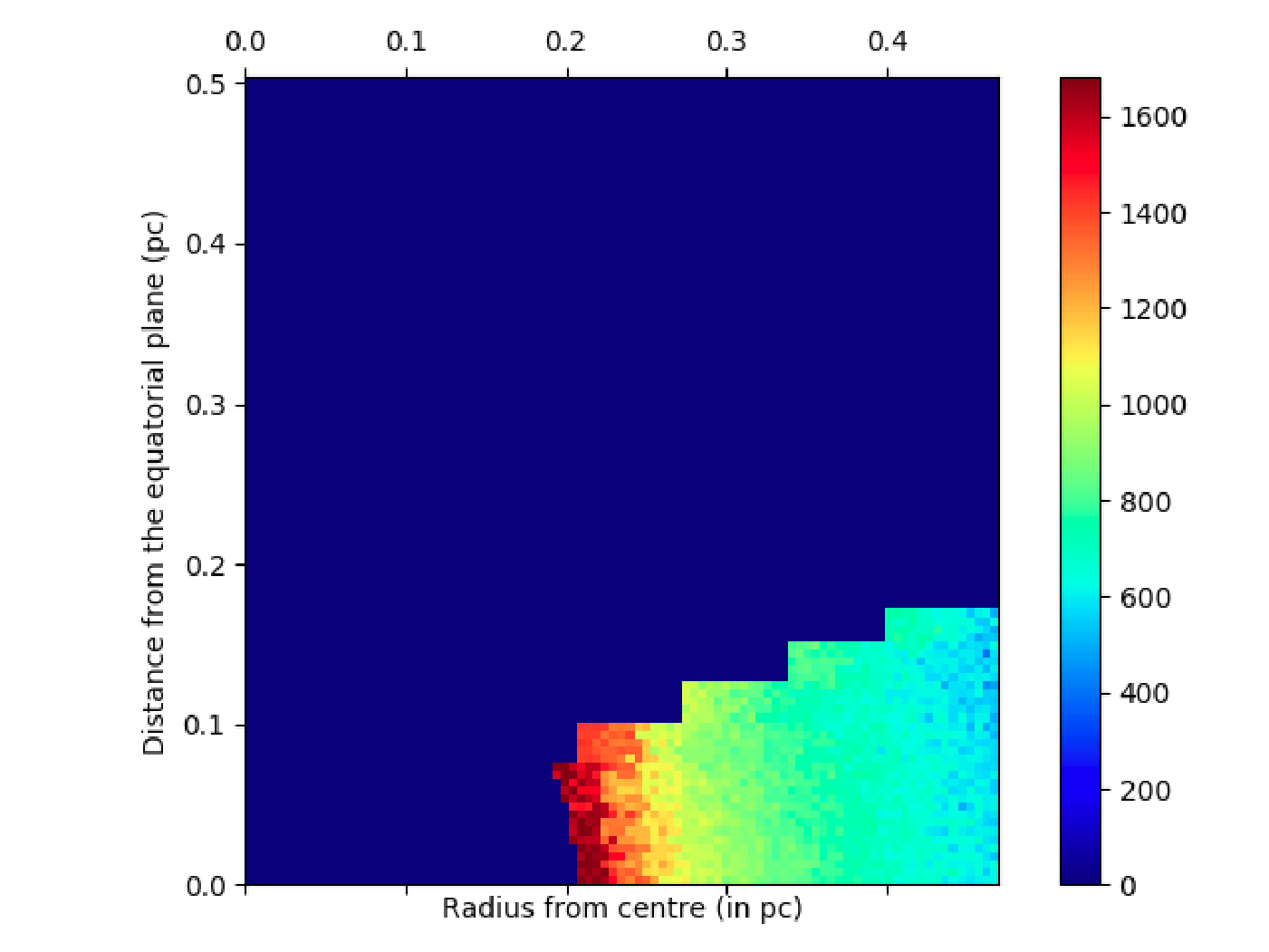}
	\caption{Temperature as a function of the radius for the MontAGN model  2.}
	\label{temp_mod2}
	\end{figure*}

    \section{Nuclear structures of NGC 1068}
    \centering
	\begin{sidewaystable*}[h]
	\centering		
	\begin{tabularx}{0.8\textwidth}{X|c|c|c|c|X}
      Object & Instrument & Size & PA & i & References \\
      \hline
      \hline
      & & & & & \\
      NLR & STIS/HST & $\geq 100\ pc$ & $33\degree$ $(\bot 123\degree)$ &  $\sim -10\degree$  $(\bot 80\degree)$ & \citet{Crenshaw2000, Das2006, Poncelet2008}  \\
      Ridge & SPHERE/VLT & $\sim 10\ pc$ & $56\degree (\bot 136\degree)$ & \textit{NA} & \citet{Gratadour2015}  \\
      \hline
      & & & & & \\
      Extended torus & SPHERE/VLT & $60 \ pc \times 20\ pc$ & $118\degree$ & ($\sim 90\degree$) & \citet{Gratadour2015}  \\
      Molecular disk & ALMA & $7 - 10\ pc$ & $112\degree$ & $33\degree - 66\degree$ & \citet{GarciaBurillo2016, Gallimore2016} \\
      $350\ K$ dust & MIDI/VLTI & $13\ pc$ & $\sim145\degree$ & \textit{NA} & \citet{LopezGonzaga2014} \\ 
      $250\ K$ dust & MIDI/VLTI &  $3\ pc$ & $100\degree - 120\degree$ & \textit{NA} & \citet{LopezGonzaga2014} \\
      \hline
      & & & & & \\
      $800\ K$ dust & MIDI/VLTI &  $1,4\ pc \times 0,4\ pc$ & $135\degree - 140\degree$ & \textit{NA} & \citet{Jaffe2004, Raban2009, Burtscher2013, LopezGonzaga2014} \\
      Counter-rotating inner disk & ALMA & $0,5\ pc \leq r \leq 1,4\ pc$ & $112\degree$ & \textit{NA} & \citet{Imanishi2018, Impellizzeri2019} \\
      S1 & VLBA & $0,8\ pc \times 0,4\ pc$ & $\sim 110\degree$ & \textit{NA} & \citet{Gallimore2004} \\
      Maser spots & VLBA & $0,65\ pc \leq r \leq 1,1\ pc$ & $135\degree$ & $90\degree$ & \citet{Greenhill1997} \\
      Parsec scale outflow & ALMA & $r \sim 0,6\ pc$ & $33\degree (\bot 123\degree)$ & \textit{NA} & \citet{Gallimore2016} \\
      \hline
      & & & & & \\
      $1600\ K$ hot dust & GRAVITY/VLTI & $r = 0,23\pm 0,03\ pc$ & $130_{-4}^{+4}$\,$\degree$ & $i=70\degree \pm 5\degree$ & \citet{GRAVITY2020} \\
      $1600\ K$ hot dust & GRAVITY/VLTI & $r = 0,21^{+0,002}_{-0,003}\ pc$ & $150_{-13}^{+8}$\,$\degree$ & $i=44^{+10}_{-10}$\,$\degree$ & This work \\
	\end{tabularx}
    \caption[Comparison of the orientations observed for various sub-structures of the inner region of NGC 1068's torus.]{\label{comp} Comparison of the orientations observed for various sub-structures of the inner region of NGC 1068's torus. \textit{NA} indicates that the estimation of the parameter is not available.}
\end{sidewaystable*}

\end{appendix}

\end{document}